%% file: main.tex
\documentclass[11pt]{article}

\input{macros}

\title{Exhaustive Search Is Hard to Beat in Constant Depth\\ for Hyperclique and Orthogonal Vectors}
\date{}
\author{Nikolai Chukhin\thanks{JetBrains Research, \url{buyolitsez1951@gmail.com}}}
\hypersetup{pdftitle={Exhaustive Search Is Hard to Beat in Constant Depth for Hyperclique and Orthogonal Vectors},pdfauthor={Nikolai Chukhin}}

\begin{document}

\hypersetup{pageanchor=false}
\begin{titlepage}
\maketitle
\begingroup
\let\small\normalsize
\begin{abstract}
\input{sections/abstract}
\end{abstract}
\endgroup
\thispagestyle{empty}
\end{titlepage}
\hypersetup{pageanchor=true}
\pagenumbering{roman}
\tableofcontents
\clearpage
\pagenumbering{arabic}

\input{sections/introduction}
\input{sections/overview}
\input{sections/preliminaries}
\input{sections/planted_monomials}
\input{sections/hypercliques}
\input{sections/growing_clique}
\input{sections/polylog_wires}
\input{sections/depth_three}
\input{sections/open_problems}
\input{sections/ai_disclosure}

\bibliographystyle{alpha}
\bibliography{refs}

\end{document}

%% file: macros.tex
\usepackage[T1]{fontenc}
\usepackage{lmodern}
\usepackage[letterpaper,margin=1in]{geometry}
\usepackage{amsmath,amssymb,amsthm}
\usepackage{microtype}
\usepackage{xcolor}
\usepackage{hyperref}
\usepackage[capitalize,noabbrev]{cleveref}
\usepackage{thm-restate}

\hypersetup{breaklinks=true,colorlinks=true,allcolors=blue!55!black}
\newtheorem{theorem}{Theorem}[section]
\newtheorem{lemma}[theorem]{Lemma}

\theoremstyle{definition}
\newtheorem{question}[theorem]{Question}
\crefname{question}{Question}{Questions}

\newcommand{\cc}[1]{\ensuremath{\mathsf{#1}}}
\newcommand{\OV}{\problemabbr{OV}}
\newcommand{\Int}{\problemabbr{Int}}
\newcommand{\Clique}{\problemname{Clique}}
\newcommand{\poly}{\operatorname{poly}}
\newcommand{\E}{\mathbb E}
\newcommand{\Prb}{\mathop{\rm Pr}}
\newcommand{\bits}{\{0,1\}}

\DeclareRobustCommand{\problemname}[1]{\textup{\textsf{#1}}}
\DeclareRobustCommand{\problemabbr}[1]{\ifmmode\mathsf{#1}\else\textnormal{\textsf{#1}}\fi}

%% file: sections/abstract.tex
A~hyperclique in~an~$r$-uniform hypergraph is~a~set of~vertices in~which every $r$-subset is~an~edge.
A~simple exhaustive search algorithm detects a~hyperclique on~$\ell$ vertices in~an~$r$-uniform hypergraph on~$m$ vertices by~checking all $\binom m\ell$ sets of~$\ell$ vertices.
This algorithm can be~expressed as~a~depth-two circuit with at~most $m^\ell$ gates and wires.
For hypercliques of~\emph{constant} size, Amano (2010) showed that this circuit is~almost optimal: for large~$r$, constant-depth circuits need $m^{\ell(1-O(\log r/r))}$ wires.
We~prove that this circuit remains hard to~beat even for hypercliques of~\emph{polynomial} size.
More precisely, circuits of~depth~$h$ need $m^{\Omega(\ell)}$ gates for $\ell\leq m^{1/h-\varepsilon}$, and $m^{\ell(1-O(\log r/r))}$ gates for $\ell=m^{o(1)}$ that tends to~infinity.

Our hyperclique bound and the method behind it also lead to~lower bounds for two central problems of~fine-grained complexity.
First, we~extend the bound $n^{k/4}$ of~Rossman (2008) for $k$-\Clique{} from cliques of~\emph{constant} size to~cliques of~\emph{polynomial} size: circuits of~depth~$h$ detecting a~clique on~$k$ vertices in~an~$n$-vertex graph need $n^{\Omega(k)}$ gates for $k\leq n^{1/(h-1)-\varepsilon}$.
Second, a~reduction of~Choudhury, Limaye, Sreenivasaiah, and Srinivasan (2026) turns our hyperclique bound into a~lower bound of~$n^{k-\varepsilon}$ gates for $k$-Orthogonal Vectors ($k$-\problemabbr{OV}) at~every fixed depth for vectors of~dimension $\log^{O(1)} n$; this proves the $k$-\problemabbr{OV}~Conjecture for constant-depth circuits in~polylogarithmic dimension.

Finally, for $2$-\problemabbr{OV}~in~logarithmic dimension, we~give a~simple proof that depth-three circuits need $\Omega(n^2)$ wires, which answers a~question of~Kane and Williams (2019).

%% file: sections/introduction.tex
\section{Introduction}\label{sec:introduction}

Fine-grained complexity aims to~explain why the textbook algorithms for many problems have resisted improvement for decades: it~connects these problems by~reductions to~a~few core problems and conjectures that the known algorithms for the core problems are essentially optimal~\cite{VW18}.
For example, the \problemname{3SUM} problem asks whether three of~$n$ given integers sum to~zero, and the conjecture that it~requires time $n^{2-o(1)}$ gives lower bounds for many problems~\cite{GO95,Pat10,AW14,AVW14,KPP16}.
Another famous example is~the Exponential Time Hypothesis (\cc{ETH})~\cite{IP01}, which asserts that $3$-\problemabbr{SAT} requires time $2^{\Omega(n)}$, and the Strong Exponential Time Hypothesis (\cc{SETH})~\cite{IP01,CIP09} asserts that, for every $\varepsilon>0$, there is~$k$ such that $k$-\problemabbr{SAT} requires time $2^{(1-\varepsilon)n}$; these hypotheses give tight lower bounds for many exponential-time and parameterized problems~\cite{CCFHJKX05,CHKX06,Mar10,LMS11,DHMTW14,MP14,CFKLMPPS15,CDLMNOPSW16,LMS18,ABHS22} and for many problems in~$\cc{P}$~\cite{Wil05,PW10,RV13,AW14,AVW14,Bri14,ABW15,BK15,BI16,BI18}; see the survey of~Vassilevska Williams~\cite{VW18}.

In~October 2026, within a~few weeks, all of~these conjectures were announced to~be~false.
Alman and Vassilevska Williams~\cite{AV26} solved \problemname{3SUM} on~$n$ integers of~polynomial size in~time $O(n^{1.9992})$.
Later that month, OpenAI~\cite{OAI:The-Exponential-Time-Hypothesis-is-False-October-25-2026} and Anthropic~\cite{Cla26} independently announced algorithms that count the satisfying assignments of~a~$k$-CNF formula on~$n$ variables in~time $2^{o(n)}$ for every fixed~$k$, thereby refuting both \cc{ETH} and \cc{SETH}.
Both refutations quickly led to~further work: simpler proofs and faster algorithms for \problemname{3SUM}~\cite{KL26,Maj26b,SY26,CIMY26}, and studies of~the new algorithms for $k$-CNF formulas and their consequences~\cite{Maj26a,Koz26}.
These refutations only underline the importance of~\emph{unconditional} lower bounds, which do~not rely on~any hypothesis.

However, unconditional lower bounds for general models of~computation are far beyond current techniques.
A~natural way to~prove them goes through Boolean circuits: every multitape Turing machine running in~time $T$ can be~simulated by~circuits of~size $O(T\log T)$~\cite{PF79}, so~a~lower bound on~circuit size implies a~lower bound on~running time.
In~particular, a~superpolynomial lower bound on~circuit size for a~function in~$\cc{NP}$ would separate $\cc{P}$ from $\cc{NP}$.
Yet not~even a~superlinear lower bound on~circuit size is~known for functions in~$\cc{NP}$~\cite{IM02,LY22,MS26}.
This gap motivates the study of~restricted models of~computation, which are weak enough to~admit strong lower bounds and yet strong enough to~express natural algorithms.

\subsection{Clique and Orthogonal Vectors}\label{sec:clique-and-ov}

This paper studies two classical problems for which, despite decades of~effort, no~known algorithm is~much faster than \emph{exhaustive search}, which tries every candidate solution.
The first is~$k$-$\Clique_n$, in~which one is~given an~$n$-vertex graph and needs to~decide whether it~contains a~clique on~$k$ vertices.
It~is~one of~Karp's~21 \cc{NP}-complete problems~\cite{Karp72} and the canonical $\cc{W}[1]$-complete problem of~parameterized complexity~\cite{CFKLMPPS15}.
The second is~$k$-\problemname{Orthogonal Vectors} ($k$-$\OV_{n,d}$), in~which one is~given $k$ lists $A_1,\ldots,A_k$, each of~$n$ binary vectors of~dimension~$d$, that~is, from $\bits^d$, and needs to~decide whether one can choose a~vector from each list so~that the coordinatewise product of~the chosen vectors is~zero.
Exhaustive search decides $k$-\problemname{Clique} by~checking all $\binom nk$ sets of~$k$ vertices, and $k$-\problemabbr{OV} by~checking all $n^k$ choices of~vectors.

Both problems are central to~fine-grained complexity.
Using fast matrix multiplication, Ne\v{s}et\v{r}il and Poljak~\cite{NP85} constructed an~algorithm for $k$-\problemname{Clique} with running time $n^{\omega k/3+O(1)}$, where $\omega<3$ is~the matrix multiplication exponent.
Lower bounds for $k$-\problemname{Clique} transfer to~many problems of~parameterized complexity~\cite{CFKLMPPS15}, and the conjecture that this algorithm is~essentially optimal gives lower bounds for many problems~\cite{ABV18,BW17,Cha19,DVV21,DV22}.
The $k$-\problemabbr{OV}~Conjecture asserts that exhaustive search is~essentially optimal for $k$-\problemabbr{OV}~already in~logarithmic dimension: for every fixed $k\geq2$ and every $\varepsilon>0$, there is~a~constant~$c$ such that $k$-\problemabbr{OV}~in~dimension $d=c\log n$ cannot be~solved in~time $n^{k-\varepsilon}$~\cite{Wil05,VW18}.
This conjecture gives lower bounds for many problems~\cite{Bri14,ABW15,BI18,FKR24,VW18}.
The $k$-\problemabbr{OV}~Conjecture also implies circuit lower bounds stronger than the best known~\cite{Wil24,AL25}.
To~date, the conjecture remains open even for $k=2$, and algorithms improve on~exhaustive search only slightly: for $2$-\problemabbr{OV} in~dimension $d=c\log n$, the best known algorithm, due to~Abboud, Williams, and Yu~\cite{AWY15}, runs in~time $n^{2-1/O(\log c)}$.
Until recently, the hardness of~both problems was in~turn supported by~\cc{ETH} and \cc{SETH}: an~algorithm deciding $k$-\problemname{Clique} in~time $f(k)n^{o(k)}$ would refute \cc{ETH}~\cite{CFKLMPPS15}, and an~algorithm deciding $k$-\problemabbr{OV}~in~time $n^{k-\varepsilon}\poly(d)$ for some $\varepsilon>0$, already in~dimension $d=O(\log n)$, would refute \cc{SETH}~\cite{Wil05,IPZ01}.
These reductions go~only one way, and the new results~\cite{OAI:The-Exponential-Time-Hypothesis-is-False-October-25-2026,Cla26} are not known to~give an~$f(k)n^{o(k)}$-time algorithm for $k$-\problemname{Clique} or~an~$n^{k-\varepsilon}\poly(d)$-time algorithm for $k$-\problemabbr{OV}.

This makes $k$-\problemname{Clique} and $k$-\problemabbr{OV} natural problems on~which to~study lower bounds, and many previous works proved such lower bounds for restricted models of~computation~\cite{Raz85,Beame90,Ros08,KW19,CS24,CLSS26}.
In~this paper, we~ask whether exhaustive search is~essentially optimal for one of~the most studied restricted models, \emph{constant-depth circuits}.
A~\emph{depth-$h$ De~Morgan circuit} is~a~directed acyclic graph whose inputs are variables, negated variables, or~constants, whose other vertices, called \emph{gates}, compute \cc{AND} or~\cc{OR} of~unbounded fan-in, and in~which every input-to-output path contains at~most $h$ gates.
We~measure the size of~a~circuit by~the number of~its gates and by~the number of~its edges, called \emph{wires}.
Constant-depth circuits are a~natural model for our question, for two reasons.
First, exhaustive search is~usually easy to~express in~this model; for example, for both of~our problems, it~is~a~circuit of~depth two or~three.
For $k$-\problemname{Clique}, whose input has one bit $x_e$ for every pair $e\in\binom{[n]}2$ of~vertices, it~is~the depth-two circuit with $\binom nk+1$ gates that checks every set of~$k$ vertices:
\[ k\text{-}\Clique_n=\bigvee_{S\in\binom{[n]}k}\bigwedge_{e\in\binom S2}x_e. \]
For $k$-\problemabbr{OV}, write $a_{i,j}$ for the $j$th vector of~$A_i$ and $a_{i,j}[q]$ for its $q$th coordinate; then exhaustive search is~the depth-three circuit with $O(n^kd)$ gates
\[ k\text{-}\OV_{n,d}=\bigvee_{j_1,\ldots,j_k\in[n]}\bigwedge_{q\in[d]}\left(\neg a_{1,j_1}[q]\vee\cdots\vee\neg a_{k,j_k}[q]\right). \]
Second, constant-depth circuits are among the few models for which strong unconditional lower bounds are known~\cite{Ajt83,FSS84,Yao85,Has86,Raz87,Smolensky87,HJP95,PPZ99,Has14,Wil14,HRST17,MW20}.
For example, every depth-$h$ circuit computing the parity of~$N$ bits has $2^{\Omega(N^{1/(h-1)})}$ gates~\cite{Has86}; for decades no~explicit function was known to~require $2^{\omega(N^{1/(h-1)})}$ gates~\cite{KR26,LR26}.
Recently, OpenAI~\cite{OAI:Beyond-the-Square-Root-Exponent-for-Depth-Three-Boolean-Circuits-September-23-2026} announced a~function in~$\cc{P}$ that requires $\cc{OR}\circ\cc{AND}\circ\cc{OR}$~circuits of~size $2^{\omega(\sqrt N)}$, which was later extended to~every fixed depth $h\geq3$, first with the bound $2^{\Omega((N\log N/\log\log N)^{1/(h-1)})}$~\cite{Chu26} and then with $2^{\Omega((N\log N)^{1/(h-1)})}$~\cite{LZ26}.
In~contrast to~these exponential lower bounds, both of~our problems have constant-depth circuits of~\emph{polynomial size} for every fixed~$k$, namely the exhaustive-search circuits above.
Note that, against circuits of~\emph{arbitrary depth}, a~bound $n^{\Omega(k)}$ with $k=\omega(1)$ for $k$-\problemname{Clique}, or~for $k$-\problemabbr{OV}~in~dimension $d\leq\poly(n)$, would imply $\cc{NP}\not\subseteq\cc{P/poly}$.

\paragraph{$k$-\problemname{Clique} in~constant depth.}
The complexity of~$k$-\problemname{Clique} and related subgraph detection problems for constant-depth circuits has been studied extensively~\cite{Lyn86,Beame90,Ros08,Amano10,Ros10,ChenF16,LRR17,Ros18,KR23}.
Improving a~bound of~Lynch~\cite{Lyn86}, Beame~\cite{Beame90} proved that depth-$h$ De~Morgan circuits computing $k$-$\Clique_n$ have $n^{\Omega_h(k)}$ wires for every $k\leq\log n$ and every sufficiently large~$n$, and for every constant~$k$, Rossman~\cite{Ros08} improved this bound to~$\omega(n^{k/4})$ gates, with an~exponent that does~not depend on~the depth.
For $k>\log n$, simple consequences of~these results and of~the bounds for the majority function~\cite{Has86} give either $2^{\Omega_h(\log^2n)}$ or~$2^{\Omega_h(k^{1/(h-1)})}$ gates, and for $k=\Theta(n)$, Gamarnik, Mossel, and Zadik~\cite{GMZ23} proved the average-case bound $\exp(n^{(1/2-o(1))/(h+3)})$; all of~these bounds are far below the $\binom nk$ gates of~exhaustive search for $\log n < k = o(n)$.

\paragraph{$k$-\problemabbr{OV}~in~constant depth.}
Kane and Williams~\cite{KW19} started the study of~the \problemabbr{OV}~Conjecture in~restricted models of~computation, and they credit Ramamohan Paturi with the question whether it~holds for constant-depth circuits.
They proved the \problemabbr{OV}~Conjecture for formulas and branching programs, and they conjectured that depth-three circuits for \problemabbr{OV}~require $n^{2-\delta}$ wires for some $\delta>0$~\cite[Conjecture~10]{KW19}.
Choudhury and Sreenivasaiah~\cite{CS24} confirmed this conjecture for $\cc{OR}\circ\cc{AND}\circ\cc{OR}$~circuits whose bottom gates have constant fan-in: for $k\leq d$, such circuits for $k$-\problemabbr{OV}~need $\Omega(n^k)$ gates.\footnote{More precisely, for $\cc{OR}\circ\cc{AND}\circ\cc{OR}$~circuits whose bottom gates contain at~most $t$ negative literals, they prove the top fan-in bound $\Omega((n/t)^k)$ for $k\leq d$~\cite{CS24}.}
At~every fixed depth, Choudhury, Limaye, Sreenivasaiah, and Srinivasan~\cite{CLSS26} proved that $k$-\problemabbr{OV}~in~dimension $d=n^\varepsilon$ requires $n^{k-\varepsilon}$ wires.
In~dimension $d\geq Ck\log n$ for an~absolute constant~$C$, Lin~\cite{Lin26rev} gave a~bound $n^{\Omega(k)}$ on~the number of~gates at~every fixed depth for every sufficiently large constant~$k$.

\subsection{Hypercliques of~Polynomial Size}\label{sec:intro-hypercliques}

The bounds of~\cite{CLSS26} for $k$-\problemabbr{OV}~are proven by~a~reduction from \emph{hypercliques}: a~hyperclique in~an~$r$-uniform hypergraph is~a~set of~vertices in~which every $r$-subset is~an~edge.
Their reduction turns the problem of~detecting a~hyperclique on~$\ell$ vertices in~an~$r$-uniform hypergraph on~$m$ vertices into an~instance of~$k$-\problemabbr{OV}~with $n\leq m^{\ell/k}$ vectors of~dimension $d=m^{O(r)}$, and they combine this reduction~with a~lower bound of~Amano~\cite{Amano10}.
Amano proved that, for all \emph{constants} $\ell>r>2$, constant-depth circuits detecting a~hyperclique on~$\ell$ vertices in~an~$r$-uniform hypergraph on~$m$ vertices have $m^{\ell(1-O(\log r/r))}$ wires.
Since Amano's bound requires constant~$\ell$, this dimension is~polynomial in~$n$, because $n=\binom m{\ell/k}\geq m$.
To~reach polylogarithmic dimension, the hyperclique size $\ell$ must be~polynomial in~$m$.
Thus, through this reduction, a~near-optimal bound for $k$-\problemabbr{OV}~in~polylogarithmic dimension follows from a~near-optimal bound for detecting hypercliques of~\emph{polynomial} size.
Also, cliques are the hypercliques for $r=2$, so~both of~our problems come down to~detecting hypercliques of~polynomial size.

The lower bounds of~\cite{Ros08,Amano10,LRR17} for cliques and hypercliques build on~the same method: they plant a~clique or~a~hyperclique into a~random graph or~hypergraph whose edge probability is~at~or~slightly below the threshold at~which such cliques or~hypercliques appear, and they analyze the circuit with H{\aa}stad's switching lemma~\cite{Has86}.
For cliques and hypercliques on~a~constant number of~vertices, this threshold is~polynomially small and the random graph or~hypergraph is~sparse, but for those on~polynomially many vertices, the threshold tends to~one and the random structure is~dense.
Their analysis works only in~the sparse case, so~these bounds hold only for cliques and hypercliques of~constant size, and those of~polynomial size need a~different argument.

\subsection{Our Results}\label{sec:intro-results}

Our main contribution is~a~method for proving lower bounds for detecting cliques and hypercliques of~polynomial size: it~plants such a~clique or~hyperclique into a~dense random graph or~hypergraph and, instead of~applying the switching lemma to~the planted structure, uses the Fourier tail bound of~Tal~\cite{Tal17} for constant-depth circuits.

This method gives new lower bounds for constant-depth De~Morgan circuits detecting hypercliques, which we~then transfer to~lower bounds for $k$-\problemname{Clique} and $k$-\problemabbr{OV}.
We~start with hypercliques: let $\ell$-$\Clique^r_m$ be~the Boolean function that takes an~$r$-uniform hypergraph on~$m$ vertices, given by~one bit for every $r$-subset of~the vertices, and outputs one exactly when the hypergraph contains a~hyperclique on~$\ell$ vertices.

\begin{theorem}[Informal version of~\Cref{thm:hyperclique-explicit}]\label{thm:main-hyperclique}
	For all integers $h,r\geq2$ and every $\varepsilon>0$, and sufficiently large~$m, \ell$ where $\ell\leq m^{1/h-\varepsilon}$, every depth-$h$ circuit computing $\ell$-$\Clique^r_m$ has $m^{\Omega_{h,r,\varepsilon}(\ell)}$ gates.
	Moreover, if~$\ell$ tends to~infinity and $\ell=m^{o(1)}$, then every depth-$h$ circuit computing $\ell$-$\Clique^r_m$ has $m^{(c_r-o(1))\ell}$ gates, where $c_r=(1-1/r)r^{-1/(r-1)}$.
\end{theorem}
Note that the simple exhaustive-search depth-two circuit for $\ell$-$\Clique^r_m$ has $\binom m\ell+1\leq m^\ell$ gates.
Since $c_r$ tends to~one as~$r$ tends to~infinity, for large~$r$ the second bound of~\Cref{thm:main-hyperclique} is~close to~this upper bound.
Amano~\cite{Amano10} proved bounds of~this form for hypercliques of~constant size and for the number of~wires.

For graphs, that~is, for $r=2$, \Cref{thm:main-hyperclique} gives $n^{\Omega(k)}$ gates for $k$-$\Clique_n$ when $k\leq n^{1/h-\varepsilon}$.
With one more idea, which saves one layer of~the circuit, we~extend this bound to~all $3\leq k\leq n^{1/(h-1)-\varepsilon}$.

\begin{theorem}\label{thm:main-graph-clique}
	For every integer $h\geq2$ and every $\varepsilon>0$, if~$n$ is~sufficiently large and $3\leq k\leq n^{1/(h-1)-\varepsilon}$, then every depth-$h$ circuit computing $k$-$\Clique_n$ has $n^{\Omega_{h,\varepsilon}(k)}$ gates.
	Moreover, if~$k$ tends to~infinity and $k=n^{o(1)}$, then every depth-$h$ circuit computing $k$-$\Clique_n$ has $n^{(1/4-o(1))k}$ gates.
\end{theorem}
The second claim of~\Cref{thm:main-graph-clique} for large~$k$ matches the exponent $k/4$ that Rossman proved for constant~$k$~\cite{Ros08}.
Note that some upper bound on~$k$ is~necessary, since for $k=n^{1-o(1)}$ the naive depth-two circuit for $k$-\Clique{} has only $n^{o(k)}$ gates.

Combining \Cref{thm:main-hyperclique} for large~$r$ with the reduction of~Choudhury, Limaye, Sreenivasaiah, and Srinivasan~\cite{CLSS26} gives the following lower bound for $k$-\problemabbr{OV}.

\begin{theorem}\label{thm:main-fixed-depth}
	For every integer $h\geq1$, every $\varepsilon>0$, and every fixed $k\geq2$, if~$n$ is~sufficiently large, then every depth-$h$ circuit computing $k$-$\OV_{n,d}$ in~dimension $d\geq(\log n)^{O_{h,\varepsilon}(1)}$ has at~least $n^{k(1-\varepsilon)}$ gates.
	In~particular, every such circuit in~dimension $d\geq(\log n)^{O_{h,k,\varepsilon}(1)}$ has at~least $n^{k-\varepsilon}$ gates.
\end{theorem}

\paragraph{Application to~Dominating Set.}
In~the $k$-\problemname{Dominating Set} problem, one is~given an~$n$-vertex graph and is~asked whether it~has a~set of~$k$ vertices such that every vertex either belongs to~the set or~is~adjacent to~a~vertex of~the set.
It~is~the canonical $\cc{W}[2]$-complete problem~\cite{CFKLMPPS15}, it~can be~solved in~time $n^{k+o(1)}$ for every $k\geq7$~\cite{EG04,Coppersmith97,PW10}, and P\u{a}tra\c{s}cu and Williams~\cite{PW10} showed that an~$O(n^{k-\varepsilon})$-time algorithm for any $k\geq3$ and $\varepsilon>0$ would refute \cc{SETH}.
Since \cc{SETH} has been announced to~be~false~\cite{OAI:The-Exponential-Time-Hypothesis-is-False-October-25-2026,Cla26}, we~note that, as~observed by~Fischer, K\"{u}nnemann, and Redzic~\cite{FKR24}, such an~algorithm would also refute the $k$-\problemabbr{OV}~Conjecture, which remains open.
For constant-depth circuits, Lai~\cite{Lai19}, following Chen and Flum~\cite{ChenF16}, proved that $f(k)n^{o(\sqrt k)}$ gates do~not suffice even to~distinguish graphs with a~dominating set of~size~$k$ from graphs without one of~size $(\log n/\log\log n)^{1/\binom k2}$.
The reduction from $k$-\problemabbr{OV}~can be~combined with \Cref{thm:main-fixed-depth}, and it~gives an~unconditional lower bound for constant-depth circuits.

\begin{restatable}{corollary}{mainDominatingSet}\label{cor:dominating-set}
	For all $h\geq1$, $k\geq2$ and every $\delta>0$, if~$n$ is~sufficiently large, then every depth-$h$ circuit that decides whether an~$n$-vertex graph has a~dominating set of~size at~most~$k$ has at~least $n^{k-\delta}$ gates.
\end{restatable}

The reduction needs the lower bound for $k$-\problemabbr{OV} only in~polynomial dimension, so~applied to~the bounds of~Choudhury, Limaye, Sreenivasaiah, and Srinivasan~\cite{CLSS26}, it~already gives the same exponent for the number of~\emph{wires}.

\paragraph{$2$-\problemabbr{OV}~in~depth three.}
In~depth three, we~give a~simple argument that proves the conjecture of~Kane and Williams~\cite[Conjecture~10]{KW19} for $2$-\problemabbr{OV}~in~logarithmic dimension.

\begin{restatable}{theorem}{mainDepthThree}\label{thm:main-depth-three}
	For every fixed $\eta>0$, there is~a~constant $c_\eta>0$ such that, for all sufficiently large~$n$ and every $d\geq(1+\eta)\log n$, every depth-three circuit computing $2$-$\OV_{n,d}$ has at~least $c_\eta n^2$ wires.
\end{restatable}

Our proof uses a~random restriction that keeps whole rows free, so~that the restricted function is~$2$-\problemabbr{OV}~on~the free rows and every bottom gate keeps only $O(\log n)$ free literals in~expectation; the restricted circuit then translates into a~small formula, and Nechiporuk's lower bound for formulas~\cite{Nec66} applies to~it.

\paragraph{Organization.}
\Cref{sec:overview} explains the main ideas behind~\Cref{thm:main-hyperclique,thm:main-graph-clique,thm:main-fixed-depth}.
\Cref{sec:preliminaries} gives the preliminaries.
\Cref{sec:planted-monomials} develops our main technique, \Cref{sec:hypercliques,sec:growing-clique} apply it~to~hypercliques and to~cliques, and \Cref{sec:arbitrary-depth} derives the bound for \problemabbr{OV}.
\Cref{sec:depth-three} proves our result for depth three, and \Cref{sec:open-problems} discusses open problems.

\paragraph{Use of~AI~tools.}
We~used AI~tools substantially; \Cref{sec:ai-methodologies} describes how.

%% file: sections/overview.tex
\section{Overview of~Our Techniques}\label{sec:overview}

As~explained in~\Cref{sec:introduction}, our lower bounds for De~Morgan circuits of~every fixed depth come from lower bounds for detecting cliques and hypercliques.
We~prove these bounds by~a~Fourier-analytic version of~the method of~Rossman~\cite{Ros08}, in~the form of~Li, Razborov, and Rossman~\cite{LRR17}.
Their method~\cite{Ros08,LRR17} plants a~copy of~the clique in~a~sparse random graph, then finds a~subcircuit that is~sensitive to~a~large part of~the planted copy, and bounds the probability of~this event for every fixed subcircuit with H{\aa}stad's switching lemma~\cite{Has86}.
Amano~\cite{Amano10} extended Rossman's method to~detecting other patterns, including hypercliques.

Our method differs from those of~\cite{Ros08,Amano10,LRR17} in~three ways.
First, we~plant into a~random hypergraph whose edge probability may tend to~one, rather than into a~sparse one.
Second, the property of~a~gate that we~trace is~algebraic: its multilinear polynomial has a~monomial touching many planted vertices, rather than the gate being sensitive to~many planted edges.
Third, and most importantly, we~bound the probability that a~fixed gate spans a~big monomial using Fourier analysis and Tal's tail bound~\cite{Tal17}.
Together, this allows the edge probability to~tend to~one, which is~the regime of~cliques and hypercliques of~polynomial size.

Below, we~describe the main ideas in~greater detail.
We~first describe the method for hypercliques, where it~is~cleanest, and then explain the changes needed for graph cliques and for \problemabbr{OV}.
The result for depth three uses a~different, more combinatorial argument, which we~sketch in~\Cref{sec:overview-depth-three}.

\subsection{Planting a~Hyperclique and Tracing a~Monomial}\label{sec:overview-arbitrary-depth}

Recall that $\ell$-$\Clique^r_m$ receives one bit~$x_e$ for every $e$ in~the set $\binom{[m]}r$ of~all $r$-subsets of~$[m]$ and accepts exactly when the hypergraph contains a~hyperclique on~$\ell$ vertices.
Suppose that a~depth-$h$ circuit~$C$ with $G(C)$ gates computes $\ell$-$\Clique^r_m$.

The high-level idea is~the following.
Every input of~$C$ reads one edge, so~it~depends on~only $r$ vertices, while the output depends on~all $\ell$ vertices of~a~hyperclique.
Hence, for every threshold~$s$ with $r\leq s<\ell$, somewhere inside the circuit two pieces of~information, each about at~most $s$ vertices of~a~planted hyperclique, are combined into one piece about more than $s$ of~them.
We~show that every fixed gate performs such a~combination only with small probability, because it~does not know where the hyperclique is~planted, so~this gives the lower bound on~the number of~gates in~the circuit.
We~take $s=\Theta(\ell)$, a~constant fraction of~$\ell$ chosen in~Step~4.

\paragraph{Step~1: plant a~witness.}
Choose a~random $r$-uniform hypergraph~$G$ on~$[m]$ in~which every $e\in\binom{[m]}r$ is~an~edge independently with probability~$p$; for a~suitable~$p$, with high probability $G$ contains no~$\ell$-hyperclique.
Now choose an~independent uniformly random set $A\subseteq[m]$ of~$\ell$ vertices, where we~will plant the hyperclique.
With high probability, no~other $\ell$-hyperclique appears even after all missing edges of~$A$ are added, and every vertex of~$A$ belongs to~at~least one $r$-subset of~$A$ that is~not an~edge of~$G$.
Some edges inside~$A$ are missing in~$G$; we~keep them as~variables and fix every other edge to~its value in~$G$.
More generally, for a~hypergraph~$H$ with vertices in~$[m]$, the \emph{restriction}~$\rho_H$ keeps every edge of~$H$ that is~missing in~$G$ as~a~variable~$z_e$ and fixes every other edge to~its value in~$G$, and $g|_{\rho_H}$ denotes the resulting function of~these variables for a~function~$g$ of~the edge variables (\Cref{sec:tracing}).
The planting uses the restriction $\rho_{K_A}$, where $K_A$ is~the hypergraph on~$A$ whose edges are all $r$-subsets of~$A$.

We~identify every Boolean function with its unique multilinear real polynomial (\Cref{sec:prelim-multilinear}).
In~this form, the \cc{AND} is~the product, $x\wedge y=xy$, and the \cc{OR} is~the sum minus the product, $x\vee y=x+y-xy$ (note that $x^{2} = x$).
Hence a~monomial of~an~\cc{AND} or~an~\cc{OR} of~two functions is~a~monomial of~one of~them or~the product of~a~monomial of~each.
A~monomial $z^E=\prod_{e\in E}z_e$ corresponds to~a~set $E\subseteq\binom{[m]}r$ of~edges, and we~say that it~\emph{touches} the vertices that lie on~these edges.
We~measure how much a~monomial knows about the planted set by~the number of~vertices from~$A$ that a~monomial touches.

Conditioned on~the event that no~other $\ell$-hyperclique appears even after all missing edges of~$A$ are added, and every vertex of~$A$ belongs to~at~least one $r$-subset of~$A$ that is~not an~edge of~$G$, the restriction $\rho_{K_A}$ turns the function computed by~the circuit into
\[ \prod_{e\in M_A}z_e, \]
where $M_A=E(K_A)\setminus E(G)$ is~the set of~missing edges of~$A$.
Thus a~circuit is~forced to~compute one large monomial whose edges touch all $\ell$ planted vertices, while every input touches at~most $r$ of~them.

Consider first the exhaustive search circuit, the \cc{OR} over all sets of~$\ell$ vertices of~the \cc{AND} of~their edges.
After the restriction, the \cc{AND} gate of~the set~$A$ computes this monomial at~once, directly from the variables, and every other \cc{AND} gate is~fixed to~zero.
This circuit is~huge because it~has a~separate gate for every possible position of~the planted set.
Since a~small circuit cannot afford this, it~builds the monomial gradually, multiplying smaller monomials into larger ones.
At~some point, it~must multiply two monomials, each touching at~most $s$ vertices of~$A$, into one that touches more than $s$ of~them.
Intuitively, such a~step is~unlikely at~any fixed gate, because the gate does not know where the planted set is.

\paragraph{Step~2: trace the monomial.}
A~gate of~$C$ may have many inputs, and an~\cc{AND} gate multiplies the monomials of~all its inputs at~once.
To~see every multiplication of~two monomials separately, we~replace every gate by~a~path of~gates of~fan-in two that takes its inputs one at~a~time, as~in~\cite{LRR17}: an~\cc{AND} gate with inputs $f_1,\ldots,f_t$ becomes $(\cdots((f_1\wedge f_2)\wedge f_3)\cdots)\wedge f_t$, and similarly for an~\cc{OR} gate.
The nodes of~this path compute the \cc{AND} or~the \cc{OR} of~the first $j$ inputs of~the gate, which we~call the $j$th \emph{prefix} of~the gate.

Now, fix a~hypergraph~$G$ and a~set~$A$ as~in~Step~1, so~that the output computes the planted monomial $\prod_{e\in M_A}z_e$.
Start at~the output with this monomial and walk down: while some child has a~monomial that touches more than $s$ vertices, move to~that child.
Every input touches at~most $r\leq s$ vertices, so~the walk stops at~a~node with a~monomial $z^E$ that touches more than $s$ vertices, while all monomials of~its children touch at~most $s$ vertices.
This monomial did not come from a~child, so~it~arose by~multiplication: $E=E_1\cup E_2$ for a~monomial $z^{E_1}$ of~one child and a~monomial $z^{E_2}$ of~the other.
Let $B_1,B_2\subseteq A$ be~the sets of~vertices touched by~$E_1$ and~$E_2$, so~that $|B_1|,|B_2|\leq s<|B_1\cup B_2|$.

Every edge of~$E$ lies inside~$B_1$ or~inside~$B_2$, because it~belongs to~$E_1$ or~to~$E_2$.
So, the node behaves like ``a~hyperclique on~$B_1$ \cc{AND} a~hyperclique on~$B_2$''.
This is~the moment at~which the circuit combines two pieces of~information about the planted set.
We~say that the node \emph{spans} the hypergraph $H=K_{B_1}\cup K_{B_2}$ if, after the restriction $\rho_H$, it~has a~monomial that touches every vertex of~$H$.
The monomial~$z^E$ survives this restriction, because all its edges lie in~$H$, so~the node spans~$H$.
Thus, on~every outcome of~the planting, some prefix of~some gate of~$C$ spans $K_{B_1}\cup K_{B_2}$ for some $B_1,B_2\subseteq A$ with $|B_1|,|B_2|\leq s<|B_1\cup B_2|$.

\paragraph{Step~3: spanning is~unlikely.}
Fix one gate of~$C$ and two sets $P_1,P_2 \subseteq [\ell]$ of~positions inside~$A$ with $|P_1|,|P_2|\leq s<|P_1\cup P_2|$.
Let $B_1, B_2 \subseteq [m]$ be~the vertices at~these positions, so~that the hypergraph $H=K_{B_1}\cup K_{B_2}$ has $v=|P_1\cup P_2|$ vertices at~uniformly random places.
We~show that the probability over the random choice of~$G$ and~$A$ that some prefix of~the gate spans~$H$ is~roughly
\[m^{-v} \cdot p^{-|E(H)|}. \]
The main factor is~$m^{-v}$, the probability of~guessing the positions of~all $v$ vertices of~$H$.
The factor $p^{-|E(H)|}\geq1$ is~a~loss that comes from the edges of~$H$, as~we~explain below.
This loss is~small, because $p$ is~close to~one and $H$ has few edges: every edge of~$H$ lies inside~$B_1$ or~inside~$B_2$, so~$H$ has at~most $2\binom sr$ edges.

We~use Fourier analysis to~bound this probability.
For a~fixed prefix~$g$, we~prove, roughly,
\begin{equation}\label{eq:overview-spanning-fourier} \Prb[\text{$g$ spans $H$}]\leq p^{-|E(H)|}\sum_{S\subseteq\binom{[m]}r}N_H(S)\,\widehat g_p(S)^2, \end{equation}
where $\widehat g_p(S)$ are the $p$-biased Fourier coefficients of~$g$, and $N_H(S)$ is~the number of~edge covers of~$H$ using at~most $v$ edges from~$S\cap E(H)$ (recall that an~\emph{edge cover} of~$H$ is~a~subset of~edges whose union contains all vertices of~$H$).
Thus every set~$S$ enters with its Fourier weight $\widehat g_p(S)^2$ once for every cover of~$H$ inside~$S$.

To~bound this sum, we~split the sets~$S$ at~a~degree threshold $D\approx\lambda(\log G(C))^{h-1}\ll m$, where $\lambda$ is~a~parameter chosen below.
A~set $S$ of~at~most $D$ edges touches at~most $rD$ vertices, so, for $D$ much smaller than~$m$, it~rarely contains a~cover of~the randomly placed hypergraph~$H$, and the expectation of~$N_H(S)$ can be~bounded by~roughly $(\mathrm eD/m)^v$.
For a~set $S$ of~more than $D$ edges, we~use a~trivial bound $N_H(S)\leq m^{O(v)}$ on~the number of~all covers with at~most $v$ edges, and use Tal's tail bound for constant-depth circuits~\cite{Tal17} to~bound the total weight of~such sets by~$2^{-\lambda}$.
Taking $\lambda=O(v\log m)$ makes $2^{-\lambda}$ smaller than $m^{-O(v)}$.
So~the sets of~low degree give the main term $p^{-|E(H)|}(\mathrm eD/m)^v$, and the sets of~high degree give a~negligible term.

Applying~\eqref{eq:overview-spanning-fourier} to~each of~the $t$ prefixes $g_1,\ldots,g_t$ of~a~gate separately would multiply the main term by~the fan-in~$t$, and the union bound in~Step~4 would then count wires rather than gates.
Instead, we~apply~\eqref{eq:overview-spanning-fourier} to~the differences $g_j-g_{j-1}$.
If~$g_i$ spans~$H$, then so~does one of~the differences $g_j-g_{j-1}$ with $j\leq i$, because they sum to~$g_i-g_0$.
Moreover, the prefixes change monotonically, so~at~every input at~most one difference is~nonzero, and the Fourier weights of~all differences together sum to~at~most one, as~for a~single Boolean function.
Hence the main term appears only once per gate, and the fan-in~$t$ enters only the negligible term.

It~remains to~explain where inequality~\eqref{eq:overview-spanning-fourier} comes from.
The problem is~that our spanning argument concerns the \emph{monomial basis}, formed by~the products $z^S=\prod_{e\in S}z_e$, whereas Tal's bound~\cite{Tal17} is~based on~the coefficients of~$g$ in~the \emph{$p$-biased Fourier basis}, formed by~the products $\prod_{e\in S}(z_e-p)/\sqrt{p(1-p)}$.
Moreover, a~nonzero coefficient in~one basis does not give a~nonzero coefficient in~the other.
To~connect the two, we~note that a~spanning monomial~$z^E$ may contain many edges, but to~touch all vertices of~$H$ it~suffices to~keep, for every vertex of~$H$, one edge of~$E$ that contains~it, so~the kept edges form an~edge cover $J\subseteq E$ of~$H$.

To~isolate the monomials that contain~$J$, we~use, for every $e\in\binom{[m]}r$, the derivative $\Delta_eg=g|_{x_e=1}-g|_{x_e=0}$, which keeps exactly the monomials of~$g$ that contain the edge~$e$ and removes $e$ from them.
Applying it~for every edge of~$J$ gives $\Delta_Jg$, whose monomial $z^{E\setminus J}$ has the same nonzero coefficient as~$z^E$ in~$g$, so~$(\Delta_Jg)|_{\rho_H}$ is~not identically zero.
Then, the same derivative is~easy to~compute in~the $p$-biased basis.
Indeed, every basis function is~linear in~each~$z_e$, and $\Delta_e$ maps $(z_e-p)/\sqrt{p(1-p)}$ to~the constant $1/\sqrt{p(1-p)}$ and kills every basis function without~$z_e$.
Hence $\Delta_Jg$ consists of~the Fourier terms of~$g$ on~sets that contain~$J$, each multiplied by~$[p(1-p)]^{-|J|/2}$, and then, by~Parseval's identity:
\[ \E_p[(\Delta_Jg)^2]=[p(1-p)]^{-|J|}\sum_{S\supseteq J}\widehat g_p(S)^2. \]
Since $\Delta_Jg$ takes integer values, the probability that it~is~nonzero at~a~random point is~at~most $\E_p[(\Delta_Jg)^2]$.

Now suppose that $g$ spans~$H$ with a~monomial that contains a~cover~$J$.
Then two events occur.
First, all edges of~$J$ are absent in~$G$ (because only missing edges become variables), which happens with probability $(1-p)^{|J|}$.
Second, $(\Delta_Jg)|_{\rho_H}$ is~not identically zero.
The function $(\Delta_Jg)|_{\rho_H}$ depends only on~the missing edges of~$H$ outside~$J$, and setting them to~constants means adding some of~them to~$G$; so~$\Delta_Jg$~is~nonzero at~a~point $y\in\bits^{\binom{[m]}r}$ obtained from~$G$ by~adding some missing edges of~$H$.
For a~fixed point $y\in\bits^{\binom{[m]}r}$ with $q$ edges of~$H$ outside~$J$, the probability that all edges of~$G$ are present in~$y$ is~$p^{-q}$ times the probability that $G$ equals~$y$ there.
Hence the second event has probability at~most $p^{-(|E(H)|-|J|)}\E_p[(\Delta_Jg)^2]$.
Note that the two events are independent, because the first one depends only on~the edges of~$J$, and $\Delta_Jg$ does~not depend on~them.
Finally, let $\mathcal E_J$ be~the event that $g$ spans~$H$ with a~monomial that contains~$J$; multiplying the two bounds and using the formula for~$\E_p[(\Delta_Jg)^2]$ gives
\begin{align*}
	\Prb[\mathcal E_J]&\leq(1-p)^{|J|}\cdot p^{-(|E(H)|-|J|)}\,\E_p[(\Delta_Jg)^2]\\
	&=(1-p)^{|J|}\cdot p^{-(|E(H)|-|J|)}\cdot[p(1-p)]^{-|J|}\sum_{S\supseteq J}\widehat g_p(S)^2\\
	&=p^{-|E(H)|}\sum_{S\supseteq J}\widehat g_p(S)^2.
\end{align*}
Summing over the covers~$J$ of~$H$ with at~most $v$ edges gives~\eqref{eq:overview-spanning-fourier}, because every set~$S$ appears once for every such cover inside~$S$.

Note that if~we~differentiated along the whole monomial~$E$, the count $N_H(S)$ would include all sets of~edges of~$H$ that touch all vertices of~$H$, up~to about $2^{|E(H)|}$ of~them, where $|E(H)|$ is~of~order~$\ell^r$, so~$\lambda$, and hence~$D$, would have to~be~of~order~$\ell^r$, and sets of~that many edges easily touch all vertices of~$H$, so~the bound for sets of~low degree would fail.

For subgraphs of~constant size and~sparse $G$,~\cite{Ros08,LRR17} give a~bound of~the same form, $p^{-|E(H)|}m^{-|B_1\cup B_2|+o(1)}$.
In~contrast, our method also applies to~subgraphs of~polynomial size and to~dense~$G$.

\paragraph{Step~4: the union bound.}
By~Step~3, the probability that a~prefix of~one fixed gate spans $K_{B_1}\cup K_{B_2}$ for one fixed pair of~sets of~positions is~roughly $p^{-|E(H)|}m^{-v}$.
There are $G(C)$ gates and at~most $4^\ell$ pairs $(P_1,P_2)$, since each of~the $\ell$ positions of~$A$ lies in~$P_1$, in~$P_2$, in~both, or~in~neither, while by~Step~2 one of~these spanning events happens on~the high-probability event of~Step~1.
Hence, by~the union bound, $G(C)\cdot4^\ell\cdot p^{-|E(H)|}m^{-v}\gtrsim1$.

Now we~balance the parameters.
Every vertex of~$H$ contributes a~factor $m^{-1}$, and every edge of~$H$ contributes a~factor $p^{-1}$, so~the bound is~stronger when $H$ has many vertices but few edges, in~our case $H = K_{B_1} \cup K_{B_2}$ for $|B_1|, |B_2| \le s < |B_1 \cup B_2|$.
Setting $s=\lfloor\gamma \ell\rfloor$ bounds $m^{-v} \le m^{-\gamma \ell}$.
We~take $p$ slightly below the threshold at~which hypercliques on~$\ell$ vertices appear, so~that the planted hyperclique is~still unique.
The worst case is~when $B_1$ and~$B_2$ share all but one vertex, and even then, for a~small constant~$\gamma$, the main factor $m^{-v}$ outweighs the loss $p^{-|E(H)|}$, so~the union bound gives $m^{\Omega(\ell)}$ gates.
Choosing the best~$\gamma$ gives the constant~$c_r$ of~\Cref{thm:main-hyperclique}, which tends to~one as~$r$ grows.
Every vertex in~fact contributes a~factor of~about $D/m$ rather than $m^{-1}$, where the threshold~$D$ of~Step~3 is~about $\ell^h$, so~the bound holds for $\ell$ up~to about $m^{1/h}$ (\Cref{thm:hyperclique-explicit}).

\subsection{From Hypercliques to~Cliques}\label{sec:overview-growing-clique}

For $k$-$\problemname{Clique}_n$, we~apply the method with $r=2$ to~graphs on~$n$ vertices; Step~4 then gives the lower bound $n^{\Omega(k)}$, but only for $k$ up~to about $n^{1/h}$.
However, we~push the argument a~bit more and obtain the same lower bound for $k$ up~to~$n^{1/(h-1)-\varepsilon}$, for every fixed $\varepsilon>0$.

We~save one layer by~starting the trace at~a~child of~the output gate, so~that every tested function has depth $h-1$.
For this, some child of~the output must compute a~large monomial exactly.
For an~\cc{OR} output, this holds after planting a~unique $k$-clique as~in~Step~1: the restricted function is~then the planted monomial, every child of~the output is~at~most this monomial, and some child equals one at~the all-ones point, so~this child computes the monomial exactly.
For an~\cc{AND} output, this planting does not suffice, because every child is~only at~least the planted monomial; for example, the children may be~the single variables~$z_e$.
We~therefore plant a~complete $(k-1)$-partite graph in~which $\Theta(k)$ parts have size two, so~that the restricted function is~the \cc{OR} of~the missing edges inside these parts.
Every child of~the output is~at~least this \cc{OR}, and some child is~zero at~the all-zeros point, so~this child computes exactly the \cc{OR}, and its polynomial has a~monomial touching all vertices of~these parts (\Cref{sec:growing-clique}).

\subsection{From Hypercliques to~Orthogonal Vectors}\label{sec:overview-ov}

The reduction of~\cite{CLSS26} from $\ell$-$\problemname{Clique}^{r}_{m}$ to~$k$-\problemabbr{OV} works as~follows.
Split a~potential hyperclique into $k$ blocks of~about $\ell/k$ vertices, and let each list contain one vector for every set of~$\ell/k$ vertices.
One group of~coordinates checks that the selected sets are disjoint, and another group checks every possible $r$-edge of~their union.
A~tuple is~orthogonal exactly when the selected sets form an~$\ell$-hyperclique, so~a~circuit for $k$-\problemabbr{OV}~yields a~circuit for $\ell$-$\Clique^r_m$ of~the same depth.
The lists have $n\approx m^{\ell/k}$ vectors of~dimension $d=O(m^r)$.
If~we~take $\ell=m^\theta$, then $\log n\approx m^\theta\log m/k$, so~$d$ is~polylogarithmic in~$n$, and the hyperclique bound $m^{(1-\eta)\ell}$ becomes $n^{k(1-\varepsilon)}$.
Note that if~$\ell$ is~a~constant, then the dimension would be~polynomial in~$n$.

This also explains why we~need to~start with hypercliques, rather than cliques: for $\ell=m^\theta$ with small~$\theta$, the exponent of~\Cref{thm:main-hyperclique} is~about $c_r\ell$, and $c_r$ tends to~one as~$r$ grows, whereas for graphs, that~is, for $r=2$, we~have $c_2=1/4$, so~the same reduction would give only about $n^{k/4}$ gates.

\subsection{Orthogonal Vectors in~Depth Three}\label{sec:overview-depth-three}

For depth three, we~apply a~random restriction that leaves about half of~the rows of~each list free.
The restricted function is~then $2$-\problemabbr{OV}~on~the free rows.
A~bottom gate that reads only literals keeps, in~expectation, only $O(\log n)$ free literals, whatever its original fan-in.
Unfolding the restricted circuit into a~formula by~copying every gate once for each wire leaving~it, we~get a~formula with $O(W\log n)$ leaves in~expectation, where $W$ is~the number of~wires of~the initial depth-three circuit.
On~the other hand, every row of~the restricted function has many distinct subfunctions (\Cref{lem:central-sets}), so~Nechiporuk's bound~\cite{Nec66} gives that every formula for it~has $\Omega(n^2\log n)$ leaves, and comparing the two bounds gives $W=\Omega(n^2)$.

%% file: sections/preliminaries.tex
\section{Preliminaries}\label{sec:preliminaries}

We~write $[n]=\{1,\ldots,n\}$.
We~write $\log$ for the base two logarithm and $\ln$ for the natural logarithm.

\subsection{Orthogonal Vectors}

An~instance of~$k$-$\OV_{n,d}$ consists of~$k$ lists $A_i=(a_{i,1},\ldots,a_{i,n})$, where $i\in[k]$, of~vectors from $\bits^d$, and
\[ k\text{-}\OV_{n,d}(A_1,\ldots,A_k)=1\quad\Longleftrightarrow\quad\exists(j_1,\ldots,j_k)\in[n]^k:\ \bigwedge_{i=1}^k a_{i,j_i}=0^d. \]
We~call such a~tuple $(j_1,\ldots,j_k)$ \emph{orthogonal}.
We~also call the vectors $a_{i,j}$ the \emph{rows} of~the instance.
The complement of~$k$-\problemabbr{OV}~is~the monotone $k$-list intersection function
\[ k\text{-}\Int_{n,d}(A_1,\ldots,A_k)=1\quad\Longleftrightarrow\quad\forall(j_1,\ldots,j_k)\in[n]^k\ \exists q\in[d]:\ \prod_{i=1}^k a_{i,j_i}[q]=1. \]

\subsection{Boolean Circuits}\label{sec:prelim-circuits}

A~\emph{De~Morgan circuit} is~a~directed acyclic graph whose inputs are literals, that~is, variables and their negations, and constants, and whose gates compute \cc{AND} or~\cc{OR} of~arbitrary fan-in; its \emph{depth} is~the maximum number of~gates on~an~input-to-output path, and a~\emph{depth-$h$ circuit} is~a~De~Morgan circuit of~depth at~most~$h$.
The \emph{depth} of~a~gate is~the maximum number of~gates on~a~path from an~input to~this gate, including the gate itself.
The \emph{gate size} $G(C)$ of~a~circuit~$C$ is~the number of~its noninput gates, and its \emph{wire size} $W(C)$ is~the number of~its edges in~the underlying graph, which we~call \emph{wires}.
We~assume that every gate lies on~a~path to~the output gate, since removing all other gates changes neither the computed function nor the depth, and does not increase the gate or~wire size.

Fix an~order of~the inputs of~every gate~$\alpha$, and let $g_1,\ldots,g_t$ be~the functions computed at~these inputs.
The \emph{prefixes} of~$\alpha$ are the functions $f_{\alpha,0},\ldots,f_{\alpha,t}$ given by
\[ f_{\alpha,j}=g_1\wedge\cdots\wedge g_j\quad\text{if $\alpha$ is~an~\cc{AND} gate},\qquad f_{\alpha,j}=g_1\vee\cdots\vee g_j\quad\text{if $\alpha$ is~an~\cc{OR}~gate}, \]
where the empty conjunction $f_{\alpha,0}$ is~the constant one and the empty disjunction is~the constant zero.
Replacing every gate~$\alpha$ of~fan-in $t\geq2$ by~gates of~fan-in two that compute $f_{\alpha,2},\ldots,f_{\alpha,t}$, each from the previous prefix and the next input, gives an~equivalent circuit of~fan-in two.
Every prefix $f_{\alpha,j}$ with $j\geq1$ is~computed by~the subcircuits of~$C$ for the first $j$ inputs of~$\alpha$ together with one new gate, so~by~a~circuit with at~most $G(C)+1$ gates and depth at~most~$h$ if~$C$ has depth~$h$, or~at~most $h-1$ if~$\alpha$ is~not the output gate.

A~\emph{projection} is~a~map $\rho\colon\bits^N\to\bits^M$ whose every output coordinate is~a~literal of~the input variables or~a~constant, and it~is~a~projection from $f\colon\bits^N\to\bits$ to~$g\colon\bits^M\to\bits$ if~$f(x)=g(\rho(x))$ for every~$x$.
Substituting~$\rho$ into a~De~Morgan circuit for~$g$ gives a~circuit for~$f$ with no~larger depth, gate size, or~wire size.
Swapping the \cc{AND} and \cc{OR}~gates and negating every input literal and constant turns, by~De~Morgan's laws, a~circuit for~$f$ into a~circuit for~$\neg f$ with the same depth, gate size, and wire size.

A~\emph{De~Morgan formula} is~a~De~Morgan circuit with gates of~fan-in two in~which every gate feeds at~most one gate, and $L(F)$ is~the number of~leaves of~a~formula~$F$, that~is, of~its input occurrences.

A~\emph{subfunction} of~a~function~$f$ on~a~set~$Y$ of~its variables is~a~function obtained from~$f$ by~fixing all variables outside~$Y$ to~constants.
We~use the following bound of~Nechiporuk~\cite{Nec66} (see also~\cite[Theorem~6.16]{Juk12}).

\begin{theorem}[\cite{Nec66}]\label{thm:nechiporuk}
	Let the variables of~a~function~$f$ be~partitioned into blocks $Y_1,\ldots,Y_t$, and let $f$ have $s_i$ distinct subfunctions on~$Y_i$ for every $i\in[t]$.
	Then every formula~$F$ with gates of~fan-in two computing~$f$ satisfies
	\[ L(F)=\Omega\left(\sum_{i=1}^t\log s_i\right). \]
\end{theorem}

\subsection{Multilinear Polynomials}\label{sec:prelim-multilinear}

For a~finite set~$X$ of~variables and $E\subseteq X$, we~write $x^E=\prod_{e\in E}x_e$.
Every function $f\colon\bits^X\to\mathbb R$ agrees on~$\bits^X$ with a~unique multilinear polynomial $P_f\in\mathbb R[x_e : e\in X]$.
Indeed, the multilinear polynomial
\[ P_f = \sum_{a\in\bits^X}f(a)\prod_{e\in X}\left(a_ex_e+(1-a_e)(1-x_e)\right) \]
agrees with~$f$ on~$\bits^X$, since its product for~$a$ equals one at~$x=a$ and zero at~every other point of~$\bits^X$.
Hence the $2^{|X|}$ monomials $x^E$ span the $2^{|X|}$-dimensional space of~all functions $\bits^X\to\mathbb R$, so~they are linearly independent on~$\bits^X$, and $P_f$ is~unique.
We~say that $E\subseteq X$ is~a~\emph{monomial of~$f$} if~the coefficient of~$x^E$ in~$P_f$ is~nonzero.
Since $x_e^2=x_e$ on~$\bits$, for all Boolean functions $f$ and~$g$ we~have the following identities, in~which the product $P_fP_g$ is~made multilinear by~replacing every $x_e^2$ by~$x_e$:
\begin{equation}\label{eq:ml-identities} P_{f\wedge g}=P_fP_g,\qquad P_{f\vee g}=P_f+P_g-P_fP_g,\qquad P_{\neg f}=1-P_f. \end{equation}

\subsection{Fourier Tails}\label{sec:prelim-fourier}

Let $X$ be~a~finite set of~variables and $0<p<1$, and let $\mu_p$ be~the product measure on~$\bits^X$ in~which every coordinate equals one with probability~$p$.
For a~function $g\colon\bits^X\to\mathbb R$, we~write $\E_p[g]$ for the expectation of~$g(x)$ when $x\sim\mu_p$.
Every such function~$g$ has a~unique \emph{normalized $p$-biased Fourier expansion}
\[ g(x)=\sum_{S\subseteq X}\widehat g_p(S)\prod_{e\in S}\frac{x_e-p}{\sqrt{p(1-p)}}, \]
where $\widehat g_p  \colon 2^{X} \to \mathbb{R}$ are the \emph{$p$-biased Fourier coefficients} of~$g$, and Parseval's identity gives $\sum_S\widehat g_p(S)^2=\E_p[g^2]$.
For the uniform measure, we~write $\widehat g(S)=\widehat g_{1/2}(S)$.
For more details on~$p$-biased Fourier analysis, see~\cite[Section~8.4]{ODo14}.

We~use the following Fourier tail bound of~Tal.

\begin{theorem}[{\cite[Theorem~18]{Tal17}}]\label{thm:tal-tails}
	Let $h\geq1$, and let a~function $f\colon\bits^X\to\bits$ be~computed by~a~depth-$h$ circuit with at~most $\sigma\geq2$ gates.
	Then, for every $D\geq0$,
	\[ \sum_{|S|\geq D}\widehat f(S)^2\leq2\cdot2^{-D/O_h((\log\sigma)^{h-1})}. \]
\end{theorem}

\Cref{thm:tal-tails} bounds Fourier tails only for the uniform measure, whereas we~need them for~$\mu_p$.
Fourier concentration of~constant-depth circuits under product measures goes back to~Furst, Jackson, and Smith~\cite{FJS91}.
We~derive it~from \Cref{thm:tal-tails} using ideas of~Blais, O'Donnell, and Wimmer~\cite{BOW08}, and Feng, Yang, Yu, and Zhang~\cite{FYYZ26}.
We~include the short proof because their theorem is~stated in~a~different form.

\begin{lemma}\label{lem:biased-tails}
	Let $h\geq1$ be~an~integer, let $0<p<1$, and let a~function $f\colon\bits^X\to\bits$ be~computed by~a~depth-$h$ circuit with at~most $\sigma\geq2$ gates.
	Then, for every $D\geq0$,
	\[ \sum_{|S|\geq D}\widehat f_p(S)^2\leq4\cdot2^{-D/O_h((\log\sigma)^{h-1})}. \]
\end{lemma}
\begin{proof}
	Let $Z^0,Z^1\in\bits^X$ be~independent samples of~$\mu_p$, and for $y\in\bits^X$ let $Z^y\in\bits^X$ be~the point with coordinates $Z^{y_e}_e$.
	For fixed $Z=(Z^0,Z^1)$, every coordinate of~$Z^y$ is~a~constant, $y_e$, or~$\neg y_e$, so~the function $f^Z(y)=f(Z^y)$ is~computed by~a~depth-$h$ circuit with at~most $\sigma$ gates (\Cref{sec:prelim-circuits}), and \Cref{thm:tal-tails} gives $\sum_{|U|\geq r}\widehat{f^Z}(U)^2\leq2\cdot2^{-r/O_h((\log\sigma)^{h-1})}$ for every~$r$.
	Write $\chi_e(x)=(x_e-p)/\sqrt{p(1-p)}$, $A_e=(\chi_e(Z^1)+\chi_e(Z^0))/2$, and $B_e=(\chi_e(Z^1)-\chi_e(Z^0))/2$, so~that $\chi_e(Z^y)=A_e+(2y_e-1)B_e$.
	Since $\chi_e(Z^0)$ and~$\chi_e(Z^1)$ are independent with mean zero and variance one, we~have $\E[A_e]=\E[B_e]=0$ and $\E[A_e^2]=\E[B_e^2]=1/2$.
	Expanding $f=\sum_S\widehat f_p(S)\prod_{e\in S}\chi_e$ gives $\widehat{f^Z}(U)=\sum_{S\supseteq U}\widehat f_p(S)\prod_{e\in U}B_e\prod_{e\in S\setminus U}A_e$, and since the pairs $(A_e,B_e)$ are independent for different~$e$, the cross terms vanish in~expectation:
	\[ \E_Z\left[\widehat{f^Z}(U)^2\right]=\sum_{S\supseteq U}2^{-|S|}\widehat f_p(S)^2. \]
	Summing over all $U$ with $|U|\geq r$ for $r=\lceil D/2\rceil$ gives
	\[ \E_Z\left[\sum_{|U|\geq r}\widehat{f^Z}(U)^2\right]=\sum_S\widehat f_p(S)^2\Prb[\mathrm{Bin}(|S|,1/2)\geq r], \]
	and this probability is~at~least $1/2$ whenever $|S|\geq D$.
	Hence the left-hand side of~the claim is~at~most $2\cdot2\cdot2^{-r/O_h((\log\sigma)^{h-1})}$, which is~at~most the right-hand side, because $r\geq D/2$.
\end{proof}

%% file: sections/planted_monomials.tex
\section{Planting a~Hyperclique and Tracing a~Monomial}\label{sec:planted-monomials}

This section develops the arguments used in~the lower bounds for hypercliques (\Cref{sec:hypercliques}) and for cliques (\Cref{sec:growing-clique})~at~every fixed depth and combines them into \Cref{lem:planted-union-bound}, which both sections build~on.
The high level idea follows the line of~Rossman~\cite{Ros08} and~Li, Razborov, and Rossman~\cite{LRR17}.

The argument has four steps, see \Cref{sec:overview-arbitrary-depth} for the proof overview.
Briefly, we~plant a~hyperclique on~a~random set~$A$ of~vertices of~a~random hypergraph, so~that some gate computes a~monomial touching many vertices of~$A$.
\Cref{lem:planted-trace} traces this monomial down the circuit to~a~step at~which two monomials, each touching at~most $s$ vertices, are multiplied into one touching more than $s$ vertices, and \Cref{lem:dense-spanning} shows that a~fixed gate performs such a~step at~a~random position only with small probability.
\Cref{lem:planted-union-bound} combines the two by~a~union bound over the gates.

\subsection{Tracing a~Monomial Down a~Circuit}\label{sec:tracing}

We~introduce a~variable~$x_e$ for every $r$-subset~$e \subseteq [m]$, which we~call an~edge, and for a~set~$E$ of~edges we~write $V(E)=\bigcup_{e\in E}e$ and say that $E$ \emph{touches} the vertices of~$V(E)$.
For an~$r$-uniform hypergraph~$G$ on~$[m]$ and a~hypergraph~$H$ whose vertices lie in~$[m]$, with edge sets $E(G)$ and~$E(H)$, let $\rho_H$ be~the restriction that keeps every edge of~$H$ absent in~$G$ as~a~variable and fixes every other edge to~its value in~$G$:
\[ \rho_H(x_e)=\begin{cases}
z_e & \text{if } e\in E(H)\setminus E(G),\\
1 & \text{if } e\in E(G),\\
0 & \text{otherwise}.
\end{cases} \]
For a~function $g\colon\bits^{\binom{[m]}r}\to\mathbb R$, let $g|_{\rho_H}\colon\bits^{E(H)\setminus E(G)}\to\mathbb R$ be~the function of~the variables~$z_e$ obtained by~substituting $\rho_H(x_e)$ for every~$x_e$ in~$g$.
We~say that $g$ \emph{spans}~$H$ if~$g|_{\rho_H}$ has a~monomial~$E$ with $V(E)\supseteq V(H)$.
The hypergraph~$G$ will always be~clear from the context, so~we~omit it~from these notations.
For a~set $B\subseteq[m]$, let $K_B$ be~the hypergraph with vertex set~$B$ whose edges are all $r$-subsets of~$B$; it~has no~edges if~$|B|<r$.

The following lemma shows that if~a~node of~a~circuit of~fan-in two carries a~monomial touching more than $s$ vertices, then somewhere below this node a~monomial touching more than $s$ but at~most $2s$ vertices is~created by~multiplying two monomials, each touching at~most $s$ vertices.

\begin{lemma}\label{lem:planted-trace}
	Let $G$ be~an~$r$-uniform hypergraph on~$[m]$, let $A\subseteq[m]$, and let $s\geq r$ be~an~integer.
	Let $C$ be~a~circuit on~the variables $(x_e)_{e\in\binom{[m]}r}$.
	Suppose that some gate~$\beta$ of~$C$ computes a~function~$g$ such that $g|_{\rho_{K_A}}$ has a~monomial touching more than $s$ vertices.
	Then there are a~gate~$\alpha$ of~$C$, a~prefix $f_{\alpha,j}$ of~$\alpha$ with $j\geq1$, and sets $B_1,B_2\subseteq A$ with $|B_1|,|B_2|\leq s<|B_1\cup B_2|$ such that $f_{\alpha,j}$ spans $K_{B_1}\cup K_{B_2}$.
	Moreover, if~$\beta$ is~not the output gate, then neither is~$\alpha$.
\end{lemma}
\begin{proof}
	Replace every gate of~$C$ by~gates of~fan-in two that compute its prefixes (\Cref{sec:prelim-circuits}).
	By~\eqref{eq:ml-identities}, a~monomial of~a~gate of~fan-in at~most two is~a~monomial of~one of~its children or~the union of~a~monomial of~each child.
	In~the restricted circuit, start at~the gate~$\beta$, with a~monomial~$E^*$ touching more than $s$ vertices, and, while some child has a~monomial $E'\subseteq E^*$ touching more than $s$ vertices, move to~this child and replace $E^*$ by~$E'$.
	Every input of~the restricted circuit is~a~literal or~a~constant, so~its monomials touch at~most $r\leq s$ vertices; hence the walk stops at~a~gate, which computes a~prefix $f_{\alpha,j}$ with $j\geq1$, and there $E^*=E_1\cup E_2$ for monomials $E_1$ and~$E_2$ of~its two children with $|V(E_1)|,|V(E_2)|\leq s<|V(E^*)|$.
	Put $B_1=V(E_1)$ and $B_2=V(E_2)$.
	Every edge of~$E_i$ lies inside~$B_i$, so~$E^*$ consists of~missing edges of~$K_{B_1}\cup K_{B_2}$, and $V(E^*)=B_1\cup B_2$.
	Fixing all other missing edges of~$K_A$ to~zero does not change the coefficient of~$z^{E^*}$, and it~agrees with their values in~$G$.
	Hence $f_{\alpha,j}$ spans $K_{B_1}\cup K_{B_2}$.
\end{proof}

\subsection{Spanning Events are Unlikely}\label{sec:dense-spanning}

We~call the event that some prefix of~a~gate spans a~hypergraph~$H$ a~\emph{spanning event}; by~\Cref{lem:planted-trace}, a~gate with a~monomial touching more than $s$ planted vertices forces such an~event for $H=K_{B_1}\cup K_{B_2}$.
It~remains to~show that, for every \emph{fixed} gate, a~spanning event is~unlikely when $H$ lies at~a~uniformly random position in~a~random hypergraph, even when the edge probability is~close to~one.
Let $\mu_p$ be~the distribution of~a~random $r$-uniform hypergraph on~$[m]$ in~which every $r$-subset of~$[m]$ is~an~edge independently with probability~$p$.

\begin{lemma}\label{lem:dense-spanning}
	Let $h\geq1$, $r\geq2$, and $m>r$ be~integers, let $0<p<1$, and let $\kappa\geq r$ and $\lambda\geq1$ be~real numbers.
	Let $f_0,\ldots,f_t$ be~the prefixes of~an~\cc{AND} or~\cc{OR}~gate whose $t\leq m^\kappa$ inputs are functions of~the variables $(x_e)_{e\in\binom{[m]}r}$, and suppose that every $f_j$ with $j\geq1$ is~computed by~a~depth-$h$ circuit with at~most $m^\kappa$ gates.
	Let $H_0$ be~a~fixed $r$-uniform hypergraph with $v$ labeled vertices, where $1\leq v<m$, let $H$ be~its image under a~uniformly random injection into~$[m]$, and let $G\sim\mu_p$ be~independent of~$H$.
	Then there is~an~integer $D\geq v$ with $D=O_h(v+(\kappa\log m)^{h-1}\lambda)$ such that
	\[ \Prb[\text{$f_j$ spans $H$ for some $j\in[t]$}]\leq p^{-|E(H_0)|}\left[\left(\frac{\mathrm eD}m\right)^v+m^{\kappa+rv+2}2^{-\lambda}\right]. \]
\end{lemma}

To~see why this bound is~small, consider the two terms in~the brackets.
In~our applications, $D$ is~much smaller than~$m$, so~in~the first term every vertex of~$H_0$ contributes a~factor $\mathrm eD/m=m^{-\Omega(1)}$, while every edge of~$H_0$ contributes a~factor $p^{-1}$.
The first term is~therefore small when $H_0$ has many vertices but few edges, as~$H_0=K_{P_1}\cup K_{P_2}$ with $|P_1|,|P_2|\leq s<|P_1\cup P_2|=v$ does.

We~prove \Cref{lem:dense-spanning} at~the end of~this subsection: \Cref{lem:spectral-envelope} below bounds the spanning probability in~terms of~Fourier tails, and \Cref{lem:biased-tails} of~\Cref{sec:prelim-fourier} bounds these tails.
We~first bound the probability of~a~spanning event for a~fixed function in~terms of~its Fourier tail.
We~use the $p$-biased Fourier expansion of~\Cref{sec:prelim-fourier} with respect to~the measure~$\mu_p$ defined above.

\begin{lemma}\label{lem:spectral-envelope}
	Let $H_0$ be~a~fixed $r$-uniform hypergraph with $v$ labeled vertices, where $1\leq v<m$.
	Let $H$ be~the image of~$H_0$ under a~uniformly random injection into~$[m]$, and let $G\sim\mu_p$ be~independent of~$H$, where $0<p<1$.
	Let $f_0,\ldots,f_t$ be~the prefixes of~a~gate, and put $d_i=f_i-f_{i-1}$ for $i\in[t]$, with $p$-biased Fourier coefficients $\widehat d_{i,p}(S)$.
	Then, for every integer $D\geq v$,
	\[ \Prb[\text{$f_i$ spans $H$ for some $i\in[t]$}]\leq p^{-|E(H_0)|}\left[\left(\frac{\mathrm eD}m\right)^v+(|E(H_0)|+1)^v\sum_{i=1}^t\sum_{|S|>D}\widehat d_{i,p}(S)^2\right]. \]
\end{lemma}
\begin{proof}
	Fix the embedding of~$H$.
	For an~edge variable~$x_e$ and a~real function~$g$, let $\Delta_eg$ be~the function $g|_{x_e=1}-g|_{x_e=0}$, and for a~set~$J$ of~edge variables let $\Delta_Jg$ be~obtained by~applying $\Delta_e$ for every $e\in J$; if~$g=\sum_Sc_Sx^S$, then $\Delta_Jg=\sum_{S\supseteq J}c_Sx^{S\setminus J}$.
	Hence, applying $\Delta_J$ to~the Fourier expansion gives
	\begin{align*}
		\Delta_Jg&=\sum_S\widehat g_p(S)\,\Delta_J\prod_{e\in S}\frac{x_e-p}{\sqrt{p(1-p)}}\\
		&=[p(1-p)]^{-|J|/2}\sum_{S\supseteq J}\widehat g_p(S)\prod_{e\in S\setminus J}\frac{x_e-p}{\sqrt{p(1-p)}}.
	\end{align*}
	So~Parseval's identity gives
	\begin{equation}\label{eq:derivative-identity} \E_p[(\Delta_Jg)^2]=[p(1-p)]^{-|J|}\sum_{S\supseteq J}\widehat g_p(S)^2. \end{equation}

	The prefixes of~an~\cc{AND} gate are pointwise nonincreasing, those of~an~\cc{OR}~gate are pointwise nondecreasing, and $f_0$ is~constant, so~at~every input~$x$ the sequence $f_0(x),\ldots,f_t(x)$ changes at~most once.
	Hence at~every input, at~most one of~the values $d_1(x),\ldots,d_t(x)$ is~nonzero, and it~is~$\pm1$, so~$\sum_id_i(x)^2\leq1$; taking the expectation gives
	\begin{equation}\label{eq:increment-energy} \sum_{i=1}^t\E_p[d_i^2]\leq1. \end{equation}
	Moreover, $f_i=f_0+d_1+\cdots+d_i$ with $f_0$ constant, so~$\Delta_Jf_i=\Delta_Jd_1+\cdots+\Delta_Jd_i$ for every nonempty~$J$.

	Suppose that $f_i$ spans $H$ for some $i\in[t]$ with a~monomial~$E$.
	The variables of~$f_i|_{\rho_H}$ are the edges of~$H$ that are absent in~$G$, and $E$ touches every vertex of~$H$.
	For every vertex~$u$ of~$H$, pick an~edge of~$E$ that contains~$u$, and let $J\subseteq E$ be~the set of~picked edges; then $J$ is~an~\emph{edge cover} of~$H$, that~is, a~set of~edges of~$H$ whose union is~$V(H)$, with at~most $v$ edges, all absent in~$G$.
	The coefficient of~$z^{E\setminus J}$ in~$(\Delta_Jf_i)|_{\rho_H}$ is~the coefficient of~$z^E$ in~$f_i|_{\rho_H}$, which is~nonzero, so~$(\Delta_Jf_i)|_{\rho_H}$ is~not identically zero.
	Let $x$ be~the assignment of~$G$ to~the edges outside~$H$.
	Then $\Delta_Jf_i(x,y)\neq0$ for some assignment~$y$ to~$E(H)\setminus J$ that agrees with~$G$ on~the edges of~$H$ present in~$G$ and may set some missing edges of~$H$ to~one; thus $y_e\geq G_e$ for every $e\in E(H)\setminus J$ (where $G_e\in\bits$ indicates whether $e$ is~an~edge of~$G$).

	Now fix an~edge cover~$J$ of~$H$, and let $\mathcal E_J$ be~the event that, for some $i\in[t]$, the function $f_i$ spans $H$ with a~monomial that contains~$J$; by~the previous paragraph, the event that some $f_i$ spans~$H$ is~the union of~the events~$\mathcal E_J$ over the edge covers~$J$ of~$H$ with at~most $v$ edges.
	Let $\mathcal A_J(x)$ be~the set of~assignments~$y$ to~$E(H)\setminus J$ such that $\Delta_Jf_i(x,y)\neq0$ for some $i\in[t]$, and write $w$ for the assignment of~$G$ to~$E(H)\setminus J$.
	By~the previous paragraph, if~$\mathcal E_J$ holds, then all edges of~$J$ are absent in~$G$, and $w\leq y$ coordinatewise for some $y\in\mathcal A_J(x)$.
	For a~fixed~$x$, the two conditions are independent, and the first of~them has probability $(1-p)^{|J|}$.
	For the second one, fix a~point~$y$ with $q$ ones; the points $w\leq y$ arise by~switching off some of~these ones, and switching off $k$ of~them multiplies the $\mu_p$-measure by~$((1-p)/p)^k$, so~their total measure is
	\[ \mu_p(y)\sum_{k=0}^q\binom qk\left(\frac{1-p}p\right)^k=\mu_p(y)\,p^{-q}\leq\mu_p(y)\,p^{-(|E(H)|-|J|)}. \]
	Summing over $y\in\mathcal A_J(x)$, the second condition has probability at~most $p^{-(|E(H)|-|J|)}$ times the $\mu_p$-measure of~$\mathcal A_J(x)$.
	Since $\Delta_Jf_i=\Delta_Jd_1+\cdots+\Delta_Jd_i$ and every $\Delta_Jd_j$ takes integer values, every $y\in\mathcal A_J(x)$ satisfies $\sum_i(\Delta_Jd_i(x,y))^2\geq1$, so~the $\mu_p$-measure of~$\mathcal A_J(x)$ is~at~most $\sum_i\E_y[(\Delta_Jd_i(x,y))^2]$.
	Averaging over~$x$ and applying~\eqref{eq:derivative-identity} to~every~$d_i$, we~get
	\[ \Prb[\mathcal E_J\mid H]\leq(1-p)^{|J|}p^{-(|E(H)|-|J|)}\sum_{i=1}^t\E_p[(\Delta_Jd_i)^2]=p^{-|E(H)|}\sum_{S\supseteq J}\sum_{i=1}^t\widehat d_{i,p}(S)^2. \]

	Summing over the covers~$J$ of~$H$ with at~most $v$ edges gives
	\[ \Prb[\text{$f_i$ spans $H$ for some $i\in[t]$}\mid H]\leq p^{-|E(H)|}\sum_SN_H(S)\sum_{i=1}^t\widehat d_{i,p}(S)^2, \]
	where $N_H(S)$ is~the number of~edge covers of~$H$ with at~most $v$ edges that are contained in~$S$.

	For $|S|>D$, we~use $N_H(S)\leq(|E(H_0)|+1)^v$.
	For $|S|\leq D$, we~expand $\E[N_H(S)]$ as~the sum, over all sets $J\subseteq S$ of~at~most $v$ edges, of~the probability that $J$ is~an~edge cover of~$H$.
	There are at~most $\sum_{j=0}^v\binom Dj$ such sets~$J$, and a~fixed set~$J$ is~an~edge cover of~$H$ only if~$V(H)=V(J)$.
	The vertex set $V(H)$ is~a~uniformly random $v$-subset of~$[m]$, so~this happens with probability at~most $1/\binom mv$.
	Using $\sum_{j=0}^v\binom Dj\leq(\mathrm eD/v)^v$ for $D\geq v$ and $\binom mv\geq(m/v)^v$, we~obtain
	\[ \E[N_H(S)]\leq\frac1{\binom mv}\sum_{j=0}^v\binom Dj\leq\left(\frac{\mathrm eD}m\right)^v. \]
	By~Parseval's identity and~\eqref{eq:increment-energy}, $\sum_S\sum_i\widehat d_{i,p}(S)^2=\sum_i\E_p[d_i^2]\leq1$.

	Averaging the bound over~$H$ and combining these estimates, we~obtain the claimed inequality.
\end{proof}

We~now combine \Cref{lem:spectral-envelope} with \Cref{lem:biased-tails}.

\begin{proof}[Proof of~\Cref{lem:dense-spanning}]
	Every $f_j$ with $j\geq1$ is~computed by~a~depth-$h$ circuit with at~most $m^\kappa$ gates, so~\Cref{lem:biased-tails} with $\sigma=m^\kappa$ gives an~integer $D_0=O_h((\kappa\log m)^{h-1}\lambda)$, the same for all~$j$, such that every $f_j$ has $p$-biased Fourier weight at~most $2^{-\lambda}$ above degree~$D_0$.
	Put $D=\max\{v,D_0\}$, so~that $D\geq v$ and $D=O_h(v+(\kappa\log m)^{h-1}\lambda)$.
	The increments $d_j=f_j-f_{j-1}$ satisfy $\widehat d_{j,p}(S)^2\leq2\widehat f_{j,p}(S)^2+2\widehat f_{j-1,p}(S)^2$, so~$\sum_{j=1}^t\sum_{|S|>D}\widehat d_{j,p}(S)^2\leq4t\cdot2^{-\lambda}$.
	Hence \Cref{lem:spectral-envelope} gives the claimed bound, because $4t(|E(H_0)|+1)^v\leq m^{\kappa+rv+2}$: indeed, $t\leq m^\kappa$, $|E(H_0)|+1\leq\binom mr+1\leq m^r$, and $4\leq m^2$.
\end{proof}

\subsection{The Union Bound}\label{sec:planted-union-bound}

We~now combine \Cref{lem:planted-trace} and \Cref{lem:dense-spanning} by~a~union bound over the gates and over the positions of~the traced vertices.
The resulting \Cref{lem:planted-union-bound} bounds the probability that some gate computes a~monomial touching more than $s$ vertices of~a~hyperclique planted at~a~random position, so~that \Cref{sec:hypercliques,sec:growing-clique} only have to~choose the parameters and to~show that this event is~likely.

For a~hyperclique planted on~a~set~$A$ of~$\ell$ vertices, \Cref{lem:planted-trace} gives a~gate and sets $B_1,B_2\subseteq A$ with $|B_1|,|B_2|\leq s<|B_1\cup B_2|$ such that some prefix of~the gate spans $K_{B_1}\cup K_{B_2}$, and we~take the union bound over all such gates and pairs.
Since $A$ is~random, we~index the pairs by~their positions $P_1,P_2\subseteq[\ell]$ in~the planted hyperclique, so~that each event says that some prefix of~a~fixed gate spans the image of~the fixed hypergraph $K_{P_1}\cup K_{P_2}$ at~the random position.
\Cref{lem:dense-spanning} bounds the probability of~each such event by~$\Lambda^{|P_1\cup P_2|}p^{-|E(K_{P_1}\cup K_{P_2})|}$ plus a~negligible term, where $\Lambda$ is~an~upper bound on~the factor $\mathrm eD/m$.
To~sum these bounds over all pairs, we~first show that no~pair is~worse than a~single complete hypergraph on~$s$ vertices.
The following lemma says that adding a~vertex to~a~set of~fewer than $s$ vertices, together with its at~most $\binom{s-1}{r-1}$ new edges, does~not increase the product $\Lambda^{|P_1\cup P_2|}p^{-|E(K_{P_1}\cup K_{P_2})|}$.

\begin{lemma}\label{lem:worst-pair}
	Let $r\geq2$ and $s\geq r$ be~integers, and let $0<\Lambda\leq1$ and $0<p<1$ be~real numbers with $\ln(1/\Lambda)\geq\binom{s-1}{r-1}\ln(1/p)$.
	Then all finite sets $P_1$ and~$P_2$ with $|P_1|,|P_2|\leq s<|P_1\cup P_2|$ satisfy
	\[ \Lambda^{|P_1\cup P_2|}p^{-|E(K_{P_1}\cup K_{P_2})|}\leq\Lambda^sp^{-\binom sr}. \]
\end{lemma}
\begin{proof}
	Put $a_i=|P_i|$ and $c=|P_1\cap P_2|$.
	Since $K_{P_1}$ and~$K_{P_2}$ share exactly the edges of~$K_{P_1\cap P_2}$, the left-hand side equals $\exp(-(F(a_1)+F(a_2)-F(c)))$ for the function
	\[ F(x)=x\ln\frac1\Lambda-\binom xr\ln\frac1p, \]
	and the right-hand side equals $\exp(-F(s))$.
	The increments $F(x+1)-F(x)=\ln(1/\Lambda)-\binom x{r-1}\ln(1/p)$ are nonincreasing in~$x$, and by~the hypothesis they are nonnegative for $x\leq s-1$, so~$F$ is~nondecreasing on~$\{0,\ldots,s\}$.
	We~bound the two terms of~$F(a_1)+(F(a_2)-F(c))$ separately.
	Let $j=a_2-c$ be~the number of~vertices of~$P_2$ outside~$P_1$, so~that $|P_1\cup P_2|=a_1+j$ and $j\leq a_2\leq s$.
	Since $a_1\leq s<a_1+j$, we~have $s-j<a_1\leq s$, so~$F(a_1)\geq F(s-j)$ because $F$ is~nondecreasing on~$\{0,\ldots,s\}$.
	Since $c=a_2-j\leq s-j$ and the increments are nonincreasing, comparing the increments term by~term gives
	\begin{align*}
		F(a_2)-F(c)&=\sum_{i=0}^{j-1}\bigl(F(c+i+1)-F(c+i)\bigr)\\
				   &\geq\sum_{i=0}^{j-1}\bigl(F(s-j+i+1)-F(s-j+i)\bigr)\\
				   &=F(s)-F(s-j).
	\end{align*}
	Adding the two bounds gives $F(a_1)+F(a_2)-F(c)\geq F(s)$.
\end{proof}

We~now take the union bound.

\begin{lemma}\label{lem:planted-union-bound}
	Let $h\geq1$ and $r\geq2$ be~integers, and let $m$ be~sufficiently large.
	Let $s\geq r$ and $1\leq\ell<m$ be~integers, let $0<p<1$, and let $\kappa\geq\max\{r+1,\ell/2\}$ and $0<\Lambda\leq1$ be~real numbers such that
	\[ \kappa^h(\log m)^{h+1}\leq\Lambda m\qquad\text{and}\qquad\binom{s-1}{r-1}\ln\frac1p\leq\ln\frac1\Lambda. \]
	Let $C$ be~a~depth-$(h+1)$ circuit on~the variables $(x_e)_{e\in\binom{[m]}r}$ with $G(C)\leq m^\kappa/2$, and let $\mathcal A$ be~the set of~all gates of~$C$ at~depth at~most~$h$.
	Let $G\sim\mu_p$, and let $\phi\colon[\ell]\to[m]$ be~a~uniformly random injection independent of~$G$, put $A=\phi([\ell])$, and let $\mathcal E$ be~an~event on~which some gate in~$\mathcal A$ computes a~function~$g$ such that $g|_{\rho_{K_A}}$ has a~monomial touching more than $s$ vertices.
	Then
	\[ \Prb[\mathcal E]\leq G(C)\cdot4^\ell\cdot\Lambda^sp^{-\binom sr}+m^{-\kappa}. \]
\end{lemma}

\begin{proof}
	We~may assume $s<\ell$, since otherwise every monomial of~$g|_{\rho_{K_A}}$ touches at~most $|A|=\ell\leq s$ vertices, and $\mathcal E$ is~empty.
	Every gate has at~most $G(C)+2\binom mr+2\leq m^\kappa$ inputs, because $G(C)\leq m^\kappa/2$ and $2\binom mr+2\leq m^{r+1}/2\leq m^\kappa/2$ for $m\geq3$.
	Then every prefix $f_{\alpha,j}$ with $\alpha\in\mathcal A$ and $j\geq1$ is~computed by~a~depth-$h$ circuit with at~most $G(C)+1\leq m^\kappa$ gates.

	Now suppose that $\mathcal E$ holds.
	By~\Cref{lem:planted-trace}, some prefix $f_{\alpha,j}$ with $j\geq1$ of~a~gate $\alpha\in\mathcal A$ spans $K_{B_1}\cup K_{B_2}$ for some $B_1,B_2\subseteq A$ with $|B_1|,|B_2|\leq s<|B_1\cup B_2|$.
	The sets $P_i=\phi^{-1}(B_i)\subseteq[\ell]$ satisfy $|P_1|,|P_2|\leq s<|P_1\cup P_2|$, and $K_{B_1}\cup K_{B_2}$ is~the image under~$\phi$ of~the hypergraph $H_0=K_{P_1}\cup K_{P_2}$ with the $v=|P_1\cup P_2|$ labeled vertices $P_1\cup P_2$.
	Hence $\mathcal E$ is~contained in~the union of~at~most $G(C)\cdot4^\ell$ events, one for every gate $\alpha\in\mathcal A$ and every such pair $(P_1,P_2)$, that some prefix $f_{\alpha,j}$ with $j\geq1$ spans the image of~$K_{P_1}\cup K_{P_2}$ under~$\phi$.

	For a~fixed gate and a~fixed pair, the restriction of~$\phi$ to~$P_1\cup P_2$ is~a~uniformly random injection into~$[m]$ independent of~$G$, and $1\leq v\leq\ell<m$, so~\Cref{lem:dense-spanning}, applied with $\lambda=(4r+10)\kappa\log m$ to~the prefixes of~the gate and to~$H_0$, bounds the probability of~the event by
	\[ p^{-|E(H_0)|}\left[\left(\frac{\mathrm eD}m\right)^v+m^{\kappa+rv+2}2^{-\lambda}\right]. \]
	Since $v\leq|P_1|+|P_2|\leq2s<2\ell\leq4\kappa$, we~have $D=O_h(v+(\kappa\log m)^{h-1}\lambda)=O_{h,r}((\kappa\log m)^h)$, so~$\mathrm eD\leq\kappa^h(\log m)^{h+1}\leq\Lambda m$ for $m$ sufficiently large in~terms of~$h$ and~$r$, and \Cref{lem:worst-pair} bounds the first term by~$\Lambda^sp^{-\binom sr}$.
	For the second term, the first hypothesis gives $\Lambda\geq1/m$, so~the second one gives $p^{-\binom sr}=(p^{-\binom{s-1}{r-1}})^{s/r}\leq\Lambda^{-s/r}\leq m^s$.
	Since $|E(H_0)|\leq\binom{|P_1|}r+\binom{|P_2|}r\leq2\binom sr$, $v\leq2s$, $s<2\kappa$, and $\kappa\geq3$, the second term is~at~most
	\[ m^{2s+\kappa+2rs+2}2^{-\lambda}\leq m^{(4r+6)\kappa-(4r+10)\kappa}=m^{-4\kappa}. \]
	Summing the two bounds over the at~most $G(C)\cdot4^\ell\leq m^\kappa\cdot16^\kappa\leq m^{3\kappa}$ events gives the claim.
\end{proof}

\subsection{Uniqueness of~the Planted Hyperclique}\label{sec:unique-hyperclique}

In~\Cref{sec:hypercliques}, and in~one of~the two plantings of~\Cref{sec:growing-clique}, the planted hyperclique has to~be~the only one.

\begin{lemma}\label{lem:unique-hyperclique}
	Let $r\geq2$ be~an~integer, let $0<\delta\leq1$, let $m$ and~$\ell$ be~integers with $r\leq\ell\leq m$, where $m$ is~sufficiently large, and let $0<p\leq m^{-(1+\delta)\ell/\binom \ell r}$.
	Let $A\subseteq[m]$ be~a~set of~$\ell$ vertices, and let $G\sim\mu_p$.
	Then, with probability $1-O(m^{-\delta})$, the set~$A$ is~the only $\ell$-hyperclique of~$G\cup K_A$, and every vertex of~$A$ lies on~an~edge of~$K_A$ that is~missing in~$G$.
\end{lemma}
\begin{proof}
	An~$\ell$-set $Q\neq A$ with $j\geq1$ vertices outside~$A$ is~a~hyperclique of~$G\cup K_A$ only if~$G$ contains all $\binom \ell r-\binom{\ell-j}r$ edges of~$K_Q$ that do~not lie inside~$A$.
	The ratio $\binom{\ell-j}r/\binom \ell r$ is~zero or~the product of~the $r\geq2$ factors $(\ell-j-i)/(\ell-i)\in(0,1]$ with $0\leq i<r$, so~it~is~at~most $(\ell-j)(\ell-j-1)/(\ell(\ell-1))$; hence there are at~least $\binom \ell r\frac j\ell\left(2-\frac{j-1}{\ell-1}\right)$ such edges, and this happens with probability at~most $m^{-(1+\delta)j(2-(j-1)/(\ell-1))}$.
	There are at~most $\binom \ell j\binom mj\leq(\mathrm e^2\ell m/j^2)^j$ such sets~$Q$, so~their expected number is~at~most $\mathrm e^{j\psi(j)}$, where
	\[ \psi(x)=2+\ln\ell+\ln m-2\ln x-(1+\delta)\left(2-\frac{x-1}{\ell-1}\right)\ln m. \]
	The function~$\psi$ is~convex on~$[1,\ell]$, so~it~is~at~most $\max\{\psi(1),\psi(\ell)\}$ there.
	Since $\ell\leq m$, we~have $\psi(1)=2+\ln\ell-(1+2\delta)\ln m\leq2-\delta\ln m$, and $\psi(\ell)=2-\ln\ell-\delta\ln m\leq2-\delta\ln m$.
	Hence the expected number of~$\ell$-hypercliques of~$G\cup K_A$ other than~$A$ is~at~most $\sum_{j\geq1}(\mathrm e^2m^{-\delta})^j\leq2\mathrm e^2m^{-\delta}$.
	Finally, a~fixed vertex of~$A$ lies on~no~missing edge of~$K_A$ only if~$G$ contains all $\binom{\ell-1}{r-1}=r\binom \ell r/\ell$ edges of~$K_A$ through it, which happens with probability at~most $m^{-(1+\delta)r}\leq m^{-r}$, and there are $\ell$ such vertices, where $\ell m^{-r}\leq m^{1-r}\leq m^{-\delta}$.
	Hence the claim fails with probability at~most $2\mathrm e^2m^{-\delta}+m^{-\delta}=O(m^{-\delta})$.
\end{proof}

%% file: sections/hypercliques.tex
\section{Hypercliques at~Every Fixed Depth}\label{sec:hypercliques}

This section proves \Cref{thm:hyperclique-explicit}, the formal version of~\Cref{thm:main-hyperclique} with an~explicit exponent, which \Cref{sec:arbitrary-depth} also uses for $k$-\problemabbr{OV}.
Recall that $\ell$-$\Clique^r_m$ receives one bit $x_e$ for every $e\in\binom{[m]}r$ and accepts exactly when the corresponding $r$-uniform hypergraph contains a~hyperclique on~$\ell$ vertices.
For an~integer $r\geq2$, put
\[ c_r=\max_{\gamma\geq0}\left(\gamma-\gamma^r\right)=\left(1-\frac1r\right)r^{-1/(r-1)}. \]
In~particular, $c_2=1/4$, and since $r^{-1/(r-1)}=\mathrm e^{-\ln r/(r-1)}\geq1-\ln r/(r-1)$, we~have $c_r\geq1-(1+\ln r)/(r-1)$, so~$c_r$ tends to~one as~$r$ grows.

As~explained in~\Cref{sec:overview-arbitrary-depth}, for $\ell\leq m^{1/h-\varepsilon}$, every vertex of~the traced hypergraph contributes a~factor of~about $\ell^h/m\leq m^{-h\varepsilon}$ to~the spanning probability, while a~fraction~$\gamma$ of~the planted vertices carries only about a~$\gamma^r$ fraction of~the edges of~the planted hyperclique; the exponent of~the bound is~the best balance of~these two sides, namely $\max_{\gamma\geq0}(h\varepsilon\gamma-\gamma^r)=c_r(h\varepsilon)^{r/(r-1)}$, attained at~$\gamma=(h\varepsilon/r)^{1/(r-1)}$.

\begin{theorem}[Formal version of~\Cref{thm:main-hyperclique}]\label{thm:hyperclique-explicit}
	Let $h\geq2$ and $r\geq2$ be~integers, and let $\varepsilon,\eta>0$ be~real numbers.
	There are integers $\ell_0$ and~$m_0$ such that, for every $m\geq m_0$ and every integer~$\ell$ with $\ell_0\leq\ell\leq m^{1/h-\varepsilon}$, every depth-$h$ circuit computing $\ell$-$\Clique^r_m$ has more than
	\[ m^{(c_r(h\varepsilon)^{r/(r-1)}-\eta)\ell} \]
	gates.
\end{theorem}

The rest of~this section proves \Cref{thm:hyperclique-explicit}.
We~take the bias slightly below the threshold, so~that a~hyperclique planted at~a~random position, as~in~Step~1 of~\Cref{sec:overview-arbitrary-depth}, is~the only one by~\Cref{lem:unique-hyperclique}, and bound the probability of~this event with \Cref{lem:planted-union-bound}.

\begin{proof}[Proof of~\Cref{thm:hyperclique-explicit}]
	We~first choose the constants.
	We~may assume $\varepsilon<1/h$, since otherwise no~integer $\ell\geq2$ satisfies $\ell\leq m^{1/h-\varepsilon}$.
	Choose $0<\xi\leq h\varepsilon/2$ such that the number $b=h\varepsilon-\xi$ satisfies
	\[ c_rb^{r/(r-1)}(1+\xi)^{-1/(r-1)}\geq c_r(h\varepsilon)^{r/(r-1)}-\frac\eta4; \]
	this is~possible because the left-hand side tends to~$c_r(h\varepsilon)^{r/(r-1)}$ as~$\xi$ tends to~zero.
	Put $\gamma=(b/(r(1+\xi)))^{1/(r-1)}$; it~maximizes $b\gamma-(1+\xi)\gamma^r$, which will be~the exponent of~the main term, and $0<\gamma<1$ because $0<b<1$.
	Let $\ell_0$ and~$m_0$ be~sufficiently large in~terms of~$h$, $r$, $\varepsilon$, and~$\eta$.

	Now fix $m\geq m_0$ and an~integer $\ell$ with $\ell_0\leq\ell\leq m^{1/h-\varepsilon}$, and let $s=\lfloor\gamma\ell\rfloor$; then $r\leq s<\ell$ by~the choice of~$\ell_0$.
	We~take the bias $p=m^{-(1+\xi)\ell/\binom \ell r}$ slightly below the threshold, and put $\kappa=\ell$ and $\Lambda=m^{-b}$ for \Cref{lem:planted-union-bound}.
	Both hypotheses of~this lemma hold.
	The first one holds by~the choice of~$m_0$, because $\kappa^h=\ell^h\leq m^{1-h\varepsilon}=m^{1-b-\xi}$.
	For the second one, $\ln(1/p)=(1+\xi)r\ln m/\binom{\ell-1}{r-1}$ and $\binom{s-1}{r-1}\leq\gamma^{r-1}\binom{\ell-1}{r-1}$, so~$\binom{s-1}{r-1}\ln(1/p)\leq(1+\xi)r\gamma^{r-1}\ln m=b\ln m=\ln(1/\Lambda)$.

	Suppose that a~depth-$h$ circuit~$C$ computes $\ell$-$\Clique^r_m$ with $G(C)\leq m^{(c_r(h\varepsilon)^{r/(r-1)}-\eta)\ell}$; then $G(C)\leq m^{\ell-1}\leq m^\kappa/2$, because $c_r(h\varepsilon)^{r/(r-1)}\leq1$ and $\eta\ell\geq1$.
	We~now plant the hyperclique: choose a~uniformly random injection $\phi\colon[\ell]\to[m]$, independent of~$G\sim\mu_p$, and put $A=\phi([\ell])$.
	\Cref{lem:unique-hyperclique} with $\xi$ in~place of~$\delta$, applied to~every fixed value of~$A$, shows that, with probability $1-O(m^{-\xi})$, the set~$A$ is~the only $\ell$-hyperclique even after all its missing edges are added, and the set~$M_A$ of~these missing edges touches every vertex of~$A$.
	On~this event, the restriction~$\rho_{K_A}$ turns $\ell$-$\Clique^r_m$ into $\prod_{e\in M_A}z_e$, so~the output of~$C$ has the monomial $M_A$, which touches $\ell>s$ vertices.

	Finally, we~bound the probability of~this event.
	\Cref{lem:planted-union-bound} applies to~it~with the parameters above, because $r\leq s$, $\ell<m$, and $\kappa=\ell\geq r+1$; here $\mathcal A$ is~the set of~all gates of~$C$, because $C$ has depth at~most~$h$.
	Using $s>\gamma\ell-1$, $b\leq1$, and $\binom sr/\binom \ell r\leq(s/\ell)^r\leq\gamma^r$, we~get
	\[ \Lambda^sp^{-\binom sr}\leq m^{-b\gamma\ell+(1+\xi)\gamma^r\ell+1}. \]
	By~the choice of~$\gamma$ and~$\xi$,
	\[ b\gamma-(1+\xi)\gamma^r=\left(1-\frac1r\right)b\gamma=c_rb^{r/(r-1)}(1+\xi)^{-1/(r-1)}\geq c_r(h\varepsilon)^{r/(r-1)}-\frac\eta4. \]
	Since $G(C)\leq m^{(c_r(h\varepsilon)^{r/(r-1)}-\eta)\ell}$, $4^\ell\leq m^{\eta\ell/4}$, and $\eta\ell\geq4$, \Cref{lem:planted-union-bound} bounds the probability of~the event of~the previous paragraph by
	\[ G(C)\cdot4^\ell\cdot m^{-(c_r(h\varepsilon)^{r/(r-1)}-\eta/4)\ell+1}+m^{-\ell}\leq m^{-\eta\ell/2+1}+m^{-\ell}\leq2m^{-1}, \]
	which contradicts its probability $1-O(m^{-\xi})$ by~the choice of~$m_0$.
\end{proof}

%% file: sections/growing_clique.tex
\section{Large Cliques at~Every Fixed Depth}\label{sec:growing-clique}

This section proves \Cref{thm:main-graph-clique}.
Recall that $k$-$\Clique_n$ receives one bit $x_e$ for every pair $e\in\binom{[n]}2$ and accepts exactly when the corresponding graph contains a~clique on~$k$ vertices.
\Cref{thm:hyperclique-explicit} with $r=2$ already gives the bound $n^{\Omega(k)}$ for large $k\leq n^{1/h-\varepsilon}$.
To~reach $k\leq n^{1/(h-1)-\varepsilon}$, we~save one layer of~the circuit by~starting the trace at~a~child of~the output gate, as~explained in~\Cref{sec:overview-growing-clique}.
Our method works only for sufficiently large~$k$, where it~gives the following bound.
For small~$k$, we~use Rossman's bound~\cite{Ros08} instead.

\begin{theorem}\label{thm:graph-clique-explicit}
	For every integer $h\geq2$ and every real $\varepsilon>0$, there are integers $k_0$ and~$n_0$ and a~real $c>0$ such that, for every $n\geq n_0$ and every integer $k_0\leq k\leq n^{1/(h-1)-\varepsilon}$, every depth-$h$ circuit computing $k$-$\Clique_n$ has more than $n^{ck}$ gates.
\end{theorem}

We~prove \Cref{thm:graph-clique-explicit} by~planting a~witness such that some child of~the output gate computes a~large monomial exactly, with one planting for each of~the two possible types of~the output gate, and by~applying \Cref{lem:planted-union-bound} to~the nonoutput gates, which have depth at~most $h-1$.

\begin{proof}[Proof of~\Cref{thm:graph-clique-explicit}]
	If~$\varepsilon\geq1/(h-1)$, then no~integer $k\geq2$ satisfies $k\leq n^{1/(h-1)-\varepsilon}$, so~$k_0=2$, $n_0=1$, and $c=1$ satisfy the theorem; we~may therefore assume $\varepsilon<1/(h-1)$.
	Put $a=(h-1)\varepsilon$, so~that $0<a<1$, and every $k\leq n^{1/(h-1)-\varepsilon}$ satisfies $k^{h-1}\leq n^{1-a}$ and, since $h\geq2$, also $k\leq n^{1-a}$.

	Put $b=a/2$ and $c=a^2/48$, let $k_0$ and~$n_0$ be~sufficiently large in~terms of~$h$ and~$\varepsilon$, and fix $n\geq n_0$ and an~integer $k_0\leq k\leq n^{1/(h-1)-\varepsilon}$; in~particular, $2k\leq2n^{1-a}\leq n$.

	We~take the bias slightly below the threshold at~which $G(n,p)$ starts to~contain $k$-cliques: put $\theta=2(1+b)/(k-1)$ and $p=n^{-\theta}$, and let $G\sim G(n,p)$, that~is, $G\sim\mu_p$ with $r=2$ and $m=n$.
	Put $t=\lfloor(b/\theta+1)/2\rfloor$ and $s=2t-1$.
	Since $b/\theta=b(k-1)/(2(1+b))$ and $b<1$, we~have $b/\theta<(k-1)/2$, and $b/\theta\geq3$ because $k\geq k_0$ is~large; hence $2\leq t<(k+1)/4$ and $3\leq s\leq b/\theta<k$.

	Suppose that a~depth-$h$ circuit~$C$ computes $k$-$\Clique_n$ and has $G(C)\leq n^{ck}$ gates; in~particular, $G(C)\leq n^{k/4}$.
	So~the output of~$C$ is~an~\cc{AND} or~an~\cc{OR}~gate.

	Next, we~show that it~suffices to~find, for each type of~the output gate, a~planting whose event has probability at~least $1/2$.
	In~both plantings below, the planted set is~$A=\phi([\ell])$ for a~uniformly random injection $\phi\colon[\ell]\to[n]$, independent of~$G$, where $\ell<2k\leq n$, and $\mathcal E$ is~an~event on~which some nonoutput gate of~$C$ computes a~function~$g$ such that $g|_{\rho_{K_A}}$ has a~monomial touching more than $s$ vertices.
	We~apply \Cref{lem:planted-union-bound} to~$\mathcal E$ with $h-1$ in~place of~$h$, $r=2$, $m=n$, $\kappa=k$, and $\Lambda=n^{-b}$; every nonoutput gate of~$C$ lies in~the set~$\mathcal A$ of~this lemma, because $C$ has depth at~most~$h$.
	Since $s\leq b/\theta$, we~have $n^{-b s}p^{-\binom s2}=n^{-b s+\theta s(s-1)/2}\leq n^{-b(s+1)/2}$, so~the lemma gives
	\begin{equation}\label{eq:graph-clique-explicit-union} \Prb[\mathcal E]\leq G(C)\cdot4^\ell\cdot n^{-b(s+1)/2}+n^{-k}. \end{equation}
	Since $s+1=2t>b/\theta-1$, $b<1$, and $b^2/(4(1+b))\geq a^2/24=2c$, we~have $b(s+1)/2>2c(k-1)-1/2\geq2ck-1$, so~$G(C)n^{-b(s+1)/2}\leq n^{-ck+1}$.
	Since also $4^\ell\leq16^k$, for $n$ sufficiently large \eqref{eq:graph-clique-explicit-union} gives
	\[ \Prb[\mathcal E]\leq16^kn^{-ck+1}+n^{-k}<\frac12, \]
	so~it~suffices to~give, for each type of~the output gate, a~planting with $\Prb[\mathcal E]\geq1/2$.

	If~the output gate is~an~\cc{OR}~gate, we~plant a~unique $k$-clique, as~in~Step~1 of~\Cref{sec:overview-arbitrary-depth}: put $\ell=k$, so~that $A$ is~a~uniformly random $k$-set.
	Let $\mathcal E$ be~the event that $A$ is~the only $k$-clique of~$G\cup K_A$ and every vertex of~$A$ lies on~a~missing edge of~$A$.
	Since $p=n^{-\theta}=n^{-(1+b)k/\binom k2}$, \Cref{lem:unique-hyperclique} with $r=2$, $m=n$, $\ell=k$, and $b$ in~place of~$\delta$, applied to~every fixed value of~$A$, gives $\Prb[\mathcal E]=1-O(n^{-b})$, which is~at~least $1/2$ for $n$ sufficiently large.
	On~$\mathcal E$, the restriction~$\rho_{K_A}$ turns the target into $\prod_{e\in M_A}z_e$, where $M_A$ is~the set of~missing edges of~$A$, and $M_A$ touches all $k$ vertices of~$A$.
	Every child of~the output \cc{OR}~gate is~at~most this product on~the restricted cube, and some child equals one on~the all-ones assignment, so~this child is~a~nonoutput gate with a~monomial touching $k>s$ vertices.

	If~the output gate is~an~\cc{AND} gate, we~plant a~complete $(k-1)$-partite graph in~which $t$ parts have size two, so~that the restricted target becomes the \cc{OR} of~the $t$ missing edges inside these parts.
	Put $\ell=k-1+t$ and $u=k-1-t$; since $t<(k+1)/4$, we~have $\ell<2k$ and $u>(3k-5)/4$.
	If~the complement of~$G[A]$ contains a~matching of~$t$ edges, let $M$ be~the first such matching in~an~order determined by~the positions~$[\ell]$.
	The $t$ pairs of~$M$ and the $u$ vertices of~$A$ not covered by~$M$, each in~a~class of~its own, form $k-1$ classes.
	Let $G_{\mathrm{base}}$ be~obtained from~$G$ by~adding all edges of~$K_A$ outside~$M$; the edges of~$M$ remain absent.
	Inside~$A$, the graph $G_{\mathrm{base}}$ is~complete $(k-1)$-partite with these classes, so~a~clique of~$G_{\mathrm{base}}$ contains at~most one vertex from each class.

	We~first show that the matching~$M$ exists with high probability.
	Otherwise, a~maximal matching in~the complement of~$G[A]$ has at~most $t-1$ edges, and its unmatched vertices form a~clique of~$G[A]$ with at~least $r_0=k-t+1$ vertices.
	Since $t<(k+1)/4$, we~have $r_0-1=k-t\geq3(k-1)/4$, so~$\theta\binom{r_0}2\geq r_0(r_0-1)/(k-1)\geq3r_0/4\geq k/2$.
	Thus this event has probability at~most $\binom\ell{r_0}p^{\binom{r_0}2}\leq2^\ell n^{-k/2}\leq(4n^{-1/2})^k$.

	Next, we~show that $G_{\mathrm{base}}$ has no~$k$-clique with high probability; since such a~clique has at~most one vertex in~each of~the $k-1$ classes, it~must use vertices outside~$A$.
	Condition on~$\phi$ and on~a~graph $G[A]$ for which $M$ exists; the edges outside $\binom A2$ remain independent Bernoulli-$p$ variables.
	Thus a~$k$-clique~$Q$ of~$G_{\mathrm{base}}$ has $j\geq1$ vertices outside~$A$, and its $k-j$ vertices in~$A$ lie in~distinct classes.
	Moreover, $Q\setminus A$ is~a~clique of~$G$, and every vertex of~$Q\cap A$ is~adjacent in~$G$ to~all vertices of~$Q\setminus A$.
	We~bound the expected number of~such pairs $(Q\setminus A,Q\cap A)$ separately for small and for large~$j$.
	Put $j_0=4k/(b\log n)$; then $j_0/(k-1)\leq8/(b\log n)$, and $j_0<k/2$ for $n$ sufficiently large.

	First, let $j<j_0$.
	Exactly $j-1$ classes contain no~vertex of~$Q$, so~$Q$ contains at~least $u-j+1$ vertices of~$A$ not covered by~$M$, where $u-j+1>(3k-5)/4-k/2+1>0$.
	We~take a~union bound over the at~most $n^j$ choices of~$Q\setminus A$ and the at~most $u^{j-1}$ choices of~a~set of~$u-j+1$ such vertices in~$Q$; the $\binom j2+j(u-j+1)=j(u-(j-1)/2)$ edges required by~these choices are distinct and lie outside $\binom A2$.
	Since $\theta(k-1)=2+a$, $\theta t\leq a/4+\theta/2$, and $\theta j<\theta j_0\leq16(1+b)/(b\log n)$, for $k\geq k_0$ and $n$ sufficiently large we~have
	\[ \theta\left(u-\frac{j-1}2\right)\geq\theta(k-1-t-j)\geq2+\frac a2. \]
	Since $u\leq k\leq n^{1-a}$, this case contributes at~most
	\[ (n^{2-a}p^{u-(j-1)/2})^j\leq n^{-aj}. \]

	Second, let $j\geq j_0$, and take a~union bound over the at~most $n^j$ choices of~$Q\setminus A$, the at~most $2^{k-1}$ choices of~the $k-j$ classes that meet~$Q$, and the at~most $2^t\leq2^k$ choices of~one vertex in~each of~these classes.
	The $\binom j2+j(k-j)\geq j(k-1)/2$ edges required by~these choices are distinct and lie outside $\binom A2$, so~this case contributes at~most
	\[ 4^kn^jp^{j(k-1)/2}\leq4^kn^{j-(1+b)j}=4^kn^{-bj}\leq n^{-bj/2}, \]
	because $n^{bj/2}\geq n^{bj_0/2}=2^{2k}$.
	Hence, since $n^{-a}$ and $n^{-b/2}$ are at~most $1/2$, the probability that $G_{\mathrm{base}}$ has a~$k$-clique is~at~most $2n^{-a}+2n^{-b/2}$, uniformly over the conditioning.
	Therefore the event~$\mathcal E$ that $M$ exists and $G_{\mathrm{base}}$ has no~$k$-clique has probability at~least $1-(4n^{-1/2})^k-2n^{-a}-2n^{-b/2}$, which is~at~least $1/2$ for $n$ sufficiently large.

	Finally, we~show that on~$\mathcal E$ some child of~the output gate has a~monomial touching more than $s$ vertices.
	On~$\mathcal E$, fix every edge of~$K_A$ outside~$M$ to~one, keep the edges of~$M$ as~variables, and fix all other edges to~their values in~$G$.
	The restricted target is~$\bigvee_{e\in M}z_e$: it~is~zero at~the all-zeros assignment, because $G_{\mathrm{base}}$ has no~$k$-clique, and adding any edge of~$M$ creates a~$k$-clique consisting of~its endpoints and one vertex from every other class.
	Every child of~the output \cc{AND} gate is~one at~every nonzero assignment, and some child is~zero at~the all-zeros assignment, so~this child computes $\bigvee_{e\in M}z_e$ exactly, and its polynomial has the coefficient $(-1)^{t+1}$ on~$\prod_{e\in M}z_e$.
	Let $R$ be~the set of~missing edges of~$K_A$ outside~$M$, and let $\sum_Fc_Fz^F$ be~the polynomial of~$g|_{\rho_{K_A}}$, where $g$ is~the function computed by~this child.
	Setting $z_e=1$ for all $e\in R$ gives the restriction above, so~$\sum_{Y\subseteq R}c_{M\cup Y}=(-1)^{t+1}\neq0$.
	Hence the child has a~monomial $M\cup Y$ touching at~least $2t>s$ vertices, and in~particular, it~is~a~nonoutput gate.

	In~both cases, $\Prb[\mathcal E]\geq1/2$ contradicts the union bound above, so~every depth-$h$ circuit computing $k$-$\Clique_n$ has more than $n^{ck}$ gates.
\end{proof}

Combining \Cref{thm:graph-clique-explicit} with Rossman's bound~\cite{Ros08} for small~$k$ gives the first claim of~\Cref{thm:main-graph-clique}, and \Cref{thm:hyperclique-explicit} with $r=2$ gives the second claim.

\begin{proof}[Proof of~\Cref{thm:main-graph-clique}]
	For the first claim, we~may assume $\varepsilon<1/(h-1)$, since otherwise no~integer $k\geq3$ satisfies $k\leq n^{1/(h-1)-\varepsilon}$.
	Let $k_0$, $n_0$, and~$c$ be~given by~\Cref{thm:graph-clique-explicit}; then, for every $n\geq n_0$ and every integer $k_0\leq k\leq n^{1/(h-1)-\varepsilon}$, every depth-$h$ circuit computing $k$-$\Clique_n$ has more than $n^{ck}$ gates.
	It~remains to~treat each of~the finitely many integers $3\leq k<k_0$; for every such~$k$, every depth-$h$ circuit computing $k$-$\Clique_n$ has more than $n^{k/4}/3\geq n^{k/8}$ gates for all sufficiently large~$n$.
	This follows directly from Rossman's~\cite{Ros08} result.

	For the second claim, fix $0<\eta<1$, and let $\ell_0$ and~$m_0$ be~given by~\Cref{thm:hyperclique-explicit} with $r=2$, $(1-\eta)/h$ in~place of~$\varepsilon$, and $\eta/2$ in~place of~$\eta$.
	Since $k$-$\Clique_n$ is~$k$-$\Clique^2_n$, $c_2(1-\eta)^2-\eta/2=(1-\eta)^2/4-\eta/2\geq1/4-\eta$, and $\ell_0\leq k\leq n^{\eta/h}$ for all sufficiently large~$n$, because $k$ tends to~infinity and $k=n^{o(1)}$, every depth-$h$ circuit computing $k$-$\Clique_n$ has more than $n^{(1/4-\eta)k}$ gates for all sufficiently large~$n$.
\end{proof}

%% file: sections/polylog_wires.tex
\section{Orthogonal Vectors at~Every Fixed Depth}\label{sec:arbitrary-depth}\label{sec:fixed-depth-polylog-wires}

This section proves \Cref{thm:main-fixed-depth} by~combining \Cref{thm:hyperclique-explicit} with the reduction of~Choudhury, Limaye, Sreenivasaiah, and Srinivasan~\cite{CLSS26}.

\begin{theorem}[{\cite[proof of~Theorem~11]{CLSS26}}]\label{thm:clss-reduction}
	Let $k,r\geq2$ and~$m$ be~integers, let $\ell\leq m$ be~a~multiple of~$k$, and put
	\[ n=\binom m{\ell/k},\qquad d=\binom k2m+\binom{r+k-1}{k-1}\binom mr. \]
	Then there is~a~projection from $\ell$-$\Clique^r_m$ to~$k$-$\OV_{n,d}$.
\end{theorem}

\begin{proof}[Proof of~\Cref{thm:main-fixed-depth}]
	It~suffices to~prove the first claim for $0<\varepsilon<1$; the second claim follows by~applying the first one with $\varepsilon/k$.
	Put $h'=\max\{h,2\}$ and $\theta=\varepsilon/(32h')$, and choose an~integer $r\geq2$ with $c_r\geq1-\varepsilon/16$, which exists because $c_r\geq1-(1+\ln r)/(r-1)$.
	Let $\ell_0$ and~$m_0$ be~given by~\Cref{thm:hyperclique-explicit} for the depth~$h'$, this~$r$, $1/h'-\theta$ in~place of~$\varepsilon$, and $\eta=\varepsilon/16$, and put $K=\lceil r/\theta\rceil+1$; all of~these depend only on~$h$ and~$\varepsilon$.

	Fix $k\geq2$, and for every~$m$, put
	\[ \ell_m=k\left\lfloor\frac{m^\theta}k\right\rfloor,\qquad q_m=\frac{\ell_m}k,\qquad n_m=\binom m{q_m},\qquad d_m=\binom k2m+\binom{r+k-1}{k-1}\binom mr. \]
	For all sufficiently large~$m$, the exponent of~\Cref{thm:hyperclique-explicit} satisfies
	\[ c_r\left(1-\frac\varepsilon{32}\right)^{r/(r-1)}-\eta\geq\left(1-\frac\varepsilon{16}\right)^2-\frac\varepsilon{16}\geq1-\frac{3\varepsilon}{16}. \]
	Every depth-$h$ circuit is~a~depth-$h'$ circuit, so~by~\Cref{thm:hyperclique-explicit} and \Cref{thm:clss-reduction} for $\ell=\ell_m$, every depth-$h$ circuit~$C$ computing $k$-$\OV_{n_m,d_m}$ satisfies
	\[ G(C)>m^{(1-3\varepsilon/16)\ell_m}\geq n_m^{k(1-3\varepsilon/16)}, \]
	because $n_m\leq m^{q_m}$.
	Moreover, $m^\theta/(2k)\leq q_m\leq m/2$, so~$\log n_m\geq q_m\log(m/q_m)\geq m^\theta/(2k)$, and hence $d_m=O_{k,r}(m^r)\leq(\log n_m)^K$ for all sufficiently large~$m$.

	This gives the lower bound on~infinitely many input lengths.
	So, it~remains to~fill the gaps between consecutive lengths~$n_m$.
	For all sufficiently large~$m$, we~have $(m+1)^\theta-m^\theta<1\leq k$, so~$q_{m+1}\in\{q_m,q_m+1\}$, and since $\binom{m+1}{q_m}=\frac{m+1}{m+1-q_m}n_m$ and $\binom{m+1}{q_m+1}=\frac{m+1}{q_m+1}n_m$, we~have $n_{m+1}\leq(m+1)n_m$.
	Since $n_m$ tends to~infinity, for a~sufficiently large~$n$, the largest~$m$ with $n_m\leq n$ exists and is~large; then $n<n_{m+1}\leq(m+1)n_m$, so~$\log n=\log n_m+O(\log m)=(1+o(1))\log n_m$.
	Repeating rows to~reach exactly $n$ rows in~every list and appending all-zero coordinates to~reach any dimension $d\geq d_m$ is~a~projection that preserves the existence of~an~orthogonal tuple.
	Hence, since $d_m\leq(\log n_m)^K\leq\lceil(\log n)^K\rceil$, every depth-$h$ circuit computing $k$-$\OV_{n,d}$ with $d\geq\lceil(\log n)^K\rceil$ has more than $n_m^{k(1-3\varepsilon/16)}\geq n^{k(1-\varepsilon)}$ gates for all sufficiently large~$n$.
\end{proof}

\subsection{Application to~Dominating Set}
We~prove \Cref{cor:dominating-set} by~a~projection from $k$-\problemabbr{OV}~in~the spirit of~the reduction of~P\u{a}tra\c{s}cu and Williams from \problemabbr{CNF-SAT}~\cite{PW10}.

\mainDominatingSet*

\begin{proof}[Proof of~\Cref{cor:dominating-set}]
	Let $a_{i,j}\in\bits^d$, for $i\in[k]$ and $j\in[m]$, be~the rows of~an~instance of~$k$-$\OV_{m,d}$, and let $n\geq km+(k+1)(d+k)$.
	The graph consists of~a~clique and an~independent set.
	The clique has one vertex $s_{i,j}$ of~color~$i$ for every row and $n-km-(k+1)(d+k)$ more vertices of~color~$1$ with the same neighbors in~the independent set as~$s_{1,1}$.
	The independent set has $k+1$ twin vertices for each of~$d$ coordinate labels and $k$ color labels.
	A~twin of~coordinate label~$q$ is~adjacent to~$s_{i,j}$ exactly when $a_{i,j}[q]=0$, and a~twin of~color label~$i$ is~adjacent to~a~clique vertex exactly when this vertex has color~$i$.
	Thus every edge variable is~a~constant or~a~negated input bit $\neg a_{i,j}[q]$.
	So~if~a~depth-$h$ circuit~$C$ decides whether an~$n$-vertex graph has a~dominating set of~size at~most~$k$, then substituting these edge values into~$C$ gives a~depth-$h$ circuit with at~most $G(C)$ gates, which computes $k$-$\OV_{m,d}$ by~the following equivalence.

	If~the rows $a_{1,j_1},\ldots,a_{k,j_k}$ are orthogonal, then $\{s_{1,j_1},\ldots,s_{k,j_k}\}$ is~a~dominating set: it~meets the clique, dominates the twins of~color label~$i$ through $s_{i,j_i}$, and dominates the twins of~coordinate label~$q$ through a~selected row with a~zero in~coordinate~$q$.
	Conversely, let $Z$ be~a~dominating set with at~most $k$ vertices.
	If~no~vertex of~$Z$ were adjacent to~the twins of~some label, then all $k+1$ twins of~this label would belong to~$Z$, which is~impossible.
	This implies that $Z$ consists of~exactly one clique vertex of~every color, and replacing an~added vertex by~$s_{1,1}$ keeps $Z$ dominating, so~we~may assume $Z=\{s_{1,j_1},\ldots,s_{k,j_k}\}$.
	For the coordinate labels, it~gives that some $a_{i,j_i}[q]$ is~zero for every $q\in[d]$, so~the selected rows are orthogonal.

	It~remains to~choose the parameters.
	We~may assume $\delta<k$, since otherwise the bound $n^{k-\delta}\leq1$ is~trivial.
	Put $\varepsilon=\delta/(2k)$, and let $(\log m)^K$, with $K$ depending only on~$h$ and~$\varepsilon$, be~the dimension threshold of~\Cref{thm:main-fixed-depth}.
	For a~large~$n$, put $m=\lfloor n/(2k)\rfloor$ and $d=\lceil(\log m)^K\rceil$, so~that $km+(k+1)(d+k)\leq n$.
	Since $m\geq n/(4k)$, \Cref{thm:main-fixed-depth} gives, for all sufficiently large~$n$,
	\[ G(C)\geq m^{k(1-\varepsilon)}\geq\left(\frac n{4k}\right)^{k-\delta/2}\geq n^{k-\delta}. \qedhere \]
\end{proof}

%% file: sections/depth_three.tex
\section{Orthogonal Vectors in~Depth Three}\label{sec:depth-three}

This section gives a~simple proof of~the following theorem.

\mainDepthThree*

We~use the following lemma, which shows that fixing all vectors of~$2$-\cc{OV} but one gives many different subfunctions.

\begin{lemma}\label{lem:central-sets}
	Let $d\geq2$ and $M_d=\binom d{\lfloor d/2\rfloor}$, and let $n_1,n_2\geq1$ be~integers.
	Consider $2$-$\OV$ on~a~first list of~$n_1$ vectors and a~second list of~$n_2$ vectors from $\bits^d$.
	Fixing all its variables outside one vector of~the first list leaves at~least $\sum_{s=0}^{\min\{n_2,M_d\}}\binom{M_d}s$ distinct functions of~this vector.
\end{lemma}
\begin{proof}
	Let $T_1,\ldots,T_{M_d}$ be~the subsets of~$[d]$ of~size $\lfloor d/2\rfloor$, so~$T_s\cap([d]\setminus T_t)=\varnothing$ if~and only if~$s=t$.
	Let $x$ be~the free vector and set the other vectors of~the first list to~$\mathbf1_d$.
	For every set $S\subseteq[M_d]$ of~at~most $n_2$ indices, let the second list consist of~the vectors $T_s$ with $s\in S$, followed by~copies of~$\mathbf1_d$.
	Every vector of~the second list is~nonempty, so~only pairs with~$x$ can be~orthogonal, and the resulting function is
	\[ x\mapsto\mathbf1[x=\varnothing]\vee\bigvee_{s\in S}\mathbf1[x\cap T_s=\varnothing]. \]
	This function equals one at~the nonempty vector $x=[d]\setminus T_t$ if~and only if~$t\in S$, so~different sets~$S$ give different functions.
\end{proof}

\begin{proof}[Proof of~\Cref{thm:main-depth-three}]
	Let $C$ be~a~depth-three circuit for $2$-$\OV_{n,d+2}$, where $d = (1+\eta)\log n$, and let $W=W(C)$.
	Consider a~random restriction~$\rho$ which makes every row, independently, \emph{free} or~\emph{fixed} with probability $1/2$ each, and sets the first $d$ coordinates of~every fixed row uniformly at~random.
	To~make fixed rows orthogonal to~no~row of~the other list, it~also sets the last two coordinates of~every fixed row to~$(1,1)$, and those of~every free row of~the first and of~the second list to~$(1,0)$ and~$(0,1)$, respectively.
	Two free rows of~different lists are then orthogonal exactly when their first $d$ coordinates are, so~the restricted function is~$2$-$\OV_{n_1,n_2,d}$, two-list \problemabbr{OV}~on~the first $d$ coordinates of~the $n_1$ free rows of~the first list and the $n_2$ free rows of~the second list.

	We~first turn the restricted circuit into a~formula and bound its expected size.
	Call a~gate of~$C$ whose inputs are all literals a~\emph{bottom gate}, and for a~bottom gate~$\alpha$ let $\tau_\alpha$ be~the number of~its free literals if~$\rho$ does not fix~$\alpha$ to~a~constant, and $\tau_\alpha=0$ otherwise.
	Then, copying every bottom gate~$\alpha$ once for each of~its $\operatorname{fanout}(\alpha)$ outgoing wires and replacing every gate by~a~binary tree of~gates of~the same type turns the restricted circuit into a~De~Morgan formula~$F_\rho$ with gates of~fan-in two that computes $2$-$\OV_{n_1,n_2,d}$, and the number $L(F_\rho)$ of~its leaves satisfies
	\[ L(F_\rho)\leq W+\sum_\alpha\operatorname{fanout}(\alpha)\,\tau_\alpha, \]
	where the sum is~over the bottom gates, and $W$ bounds the number of~literals read by~the other gates.

	If~a~bottom gate~$\alpha$ reads literals on~the first $d$ coordinates of~$s$ rows, then $\tau_\alpha\leq ds$, and each of~these rows, independently, is~fixed and gives one of~these literals the value that fixes~$\alpha$ with probability at~least $1/4$, so~$\E_\rho[\tau_\alpha]\leq ds(3/4)^s=O(d)$.
	Since $\sum_\alpha\operatorname{fanout}(\alpha)\leq W$, we~obtain $\E_\rho[L(F_\rho)]=O(dW)$.

	On~the other hand, $n_1,n_2=\Theta(n)$ with constant probability.
	For such outcomes, put $n'=\min\{n_1,n_2\}=\Theta(n)$.
	In~the first $n'$ free rows of~each list, the coordinates $d-1$ and~$d$ become $(0,1)$ in~the first list and $(1,0)$ in~the second list, and every other free row gets ones exactly in~these two of~its first $d$ coordinates.
	Then no~pair of~rows of~different lists that contains one of~the other free rows is~orthogonal, and two of~the first $n'$ rows are orthogonal exactly when they are orthogonal on~the first $d-2$ coordinates, so~the resulting formula computes $2$-$\OV_{n',d-2}$.
	Let $M=\binom{d-2}{\lfloor(d-2)/2\rfloor}$; since $M\geq2^{d-2}/(d-1)$ and $d=(1+\eta)\log n$, we~have $\log M\geq(1+\eta/2)\log n$ for all sufficiently large~$n$, so~$M\geq n'$ and $\log(M/n')\geq(\eta/2)\log n$.
	By~\Cref{lem:central-sets} in~dimension $d-2$ with $n_1=n_2=n'$, fixing all variables of~this formula outside one row of~the first list leaves at~least $\binom M{n'}\geq(M/n')^{n'}$ distinct functions of~this row.
	Nechiporuk's bound (\Cref{thm:nechiporuk}), applied with the $2n'$ rows as~the blocks, therefore gives $L(F_\rho)=\Omega(n'\cdot n'\eta\log n)=\Omega_\eta(n^2\log n)$.
	Hence $\E_\rho[L(F_\rho)]=\Omega_\eta(n^2\log n)$, and comparing with $\E_\rho[L(F_\rho)]=O(dW)$ gives $W=\Omega_\eta(n^2)$.
\end{proof}

%% file: sections/open_problems.tex
\section{Open Problems}\label{sec:open-problems}

We~conclude with the questions that we~find most interesting.

\Cref{thm:main-graph-clique} gives $n^{\Omega(k)}$ gates for every $3\leq k\leq n^\delta$, and every fixed $\delta<1/(h-1)$.
For every $n^\delta\leq k\leq n/2$, it~gives $2^{\Omega(n^\delta\log n)}$ gates by~padding.
So, the question is~whether the lower bound $n^{\Omega(k)}$ persists beyond $k=n^\delta$.

\begin{question}\label{ques:clique-larger-k}
	For every fixed $h\geq3$, do~depth-$h$ circuits for $k$-$\Clique_n$ require $n^{\Omega_h(k)}$ gates for all sufficiently large~$n$ and every $3\leq k\leq n^{1/(h-1)}$?
	Is~there a~fixed $\delta>1/(h-1)$ for which they require $n^{\Omega_h(k)}$ gates for every $3\leq k\leq n^\delta$?
\end{question}

For $h\geq4$ and $\delta>2/(h-1)$, a~bound $n^{\Omega(k)}$ at~$k=n^\delta$ would be~at~least $2^{N^{1/(h-1)+\Omega(1)}}$ for the input length $N=\binom n2$, while the strongest known depth-$h$ lower bound for an~explicit function is~$2^{\Omega((N\log N)^{1/(h-1)})}$~\cite{Chu26,LZ26}.
On~the other hand, exhaustive search uses $\binom nk+1$ gates, which is~$n^{o(k)}$ when $k=n^{1-o(1)}$.

In~a~remark in~the concluding section of~his paper~\cite[Section~6]{Ros08}, Rossman states the belief that his bound $\omega(n^{k/4})$ should also hold for cliques of~superconstant size, perhaps up~to~$k\leq\log n$.
\Cref{thm:main-graph-clique} gives the exponent $(1/4-o(1))k$ for all $k=n^{o(1)}$ that tend to~infinity, but not $\omega(n^{k/4})$: the loss $n^{o(k)}$ in~the bound may be~superpolynomial in~$n$, for example when $k$ is~close to~$\log n$.
So, a~natural question is~to~extend the range of~$k$ for which depth-$h$ circuits for $k$-$\Clique_n$ require $\omega(n^{k/4})$ gates, for example to~all $k\leq\log n$.

Finally, for $2$-\problemabbr{OV}, depth-three circuits need $\Omega(n^2)$ wires in~every dimension $d\geq(1+\eta)\log n$ (\Cref{thm:main-depth-three}).
At~every fixed depth, \Cref{thm:main-fixed-depth} gives $n^{k - \varepsilon}$ gates, but in~polylogarithmic dimension whose power depends on~$\varepsilon$ and~$k$.

\begin{question}\label{ques:fixed-k-greater-depth}
	For every fixed $k\geq2$, $h\geq3$, and $\varepsilon>0$, is~there a~constant $c>0$ such that, for all sufficiently large~$n$ and every dimension $d\geq\lceil c\log n\rceil$, depth-$h$ circuits for $k$-$\OV_{n,d}$ require $n^{k-\varepsilon}$ gates?
\end{question}

%% file: sections/ai_disclosure.tex
\section{Use of~AI and the Story of~This Paper}\label{sec:ai-methodologies}

This section follows the recommendations of~the working group report~\cite{AITCS26} and~uses its template.

\begin{flushleft}
	\renewcommand{\arraystretch}{1.3}
	\begin{tabular}{@{}l@{\qquad}c@{}}
		\textbf{Activity} & \textbf{AI used} \\
		Asking the research question & No \\
		Coming up with the approach & Yes \\
		Proof development & Yes \\
		Writing and exposition & Yes \\
		Checking for bugs & Yes \\
		Other supporting tasks & Yes \\
	\end{tabular}
\end{flushleft}

\paragraph{Details (elaborating on~the points above).}
I~used OpenAI Codex, OpenAI ChatGPT Pro with the models ChatGPT~5.6 and ChatGPT Astra, and Anthropic's Claude Opus~5.5 through Claude Code.
I~was aware of~the lower bounds for \problemabbr{OV} of~\cite{KW19, CS24}, and then I~saw the paper of~Choudhury, Limaye, Sreenivasaiah, and Srinivasan~\cite{CLSS26}.
One of~their results is~a~monotone lower bound for $k$-\problemabbr{OV}; with the help of~AI, I~found a~lifting theorem of~de~Rezende and Vinyals~\cite{RV25} that strengthens this bound, and I~wrote about~it to~Tameem Choudhury and Karteek Sreenivasaiah.
Then, after a~conversation with them, I~started looking for lower bounds against constant-depth De~Morgan circuits.
Along the way, the models proposed candidate strengthenings and proofs, some of~which became results of~this paper, as~described below.
Claude Code helped me to~restructure the paper and to~rewrite its exposition.
Below, I~describe the first versions of~the results and how they became the final ones.

\paragraph{Depth three.}
The first result, found together with ChatGPT~5.6, was \Cref{thm:main-depth-three}, with a~longer proof that was hard to~follow.
Working on~this argument, I~simplified it~a~lot and strengthened it~to~the present one-page proof in~\Cref{sec:depth-three}.

\paragraph{Cliques.}
Then, we~discussed this argument with Tameem and Karteek.
The natural next goal was to~prove constant-depth lower bounds for $k$-\problemabbr{OV}, and one approach was to~prove stronger lower bounds for cliques and transfer them to~$k$-\problemabbr{OV}~through the reduction of~\cite{CLSS26}.
Later, chatting with ChatGPT Astra, I~proved \Cref{thm:main-graph-clique}; this was a~surprise, since the result seemed too good to~be~true.
Codex developed and checked the first full proof.
I~then realized that its proof can be~stated as~a~separate method (\Cref{sec:planted-monomials}), and I~used it~to~prove both \Cref{thm:main-graph-clique,thm:main-fixed-depth}.
Afterwards, I~simplified the proofs and rewrote them.

\paragraph{Orthogonal Vectors and hypercliques.}
ChatGPT Pro helped with the research behind the bound for growing hypercliques and its consequence for $k$-\problemabbr{OV}~in~polylogarithmic dimension.
The first version of~\Cref{thm:main-fixed-depth} bounded only the number of~wires.
We~extended the Fourier bound to~all prefixes of~a~gate (\Cref{lem:dense-spanning}), which turned it~into a~bound on~the number of~gates.
Codex also found the application to~\problemname{Dominating Set} (\Cref{cor:dominating-set}) and I~verified its correctness and how interesting it~is.
Finally, I~was chatting with Claude Code a~lot in~order to~obtain a~general hyperclique bound, which is~now \Cref{thm:hyperclique-explicit}.
This gave \Cref{thm:main-hyperclique}, and \Cref{thm:main-fixed-depth} became a~short application of~it~and of~the projection of~\cite{CLSS26}.

\paragraph{Simplifications.}
The arguments written by~the models were hard to~understand, so~I~tried to~trace where each of~them comes from and to~replace it~by~a~citation of~the original source.
In~this process, the paper, which at~some point had sixty~pages, became much shorter and clearer.
As~an~example, an~agent first proved \Cref{lem:biased-tails} on~its own, without citing anyone, although its argument is~well known and not new; now the lemma follows~\cite{BOW08,FYYZ26}, and its proof is~short.

I~verified the correctness and originality of~all content myself.
I~am fully responsible for the paper's~correctness and originality.

\paragraph{Acknowledgments.}
I~am very grateful to~Tameem Choudhury and Karteek Sreenivasaiah for these discussions and for their series of~papers~\cite{CS24,CLSS26}, which raised my~interest in~this problem.
I~thank Vitaly Aksenov for reading the introduction and Ivan Mihajlin for listening to~my~ideas on~this problem.